\documentclass[pdflatex,sn-basic]{sn-jnl}

\usepackage{amsmath}
\usepackage{graphicx}
\usepackage{caption}
\usepackage{subcaption}
\usepackage{bbm}

\newcommand{\boldtext}[1]{{#1}}

\jyear{2024}%

\theoremstyle{thmstyleone}%
\theoremstyle{thmstyletwo}%

\theoremstyle{thmstylethree}%

\begin{document}

\title[Epidemiological dynamics in populations structured by neighbourhoods and households]{Epidemiological dynamics in populations structured by neighbourhoods and households}

\author*[1]{\fnm{Abby} \sur{Barlow}}\email{ahb48@bath.ac.uk}
\author[1]{\fnm{Sarah} \sur{Penington}}\email{sp2355@bath.ac.uk}
\author[1]{\fnm{Ben} \sur{Adams}}\email{ba224@bath.ac.uk}

\affil*[1]{\orgdiv{Department of Mathematical Sciences}, \orgname{The University of Bath}, \orgaddress{\street{Claverton Down}, \city{Bath}, \postcode{BA2 7AY}, \country{United Kingdom}}}

\abstract{Epidemiological dynamics are affected by the spatial and demographic structure of the host population. Households and neighbourhoods are known to be important \boldtext{groupings} but little is known about the epidemiological interplay between them. \boldtext{In order to explore the implications for infectious disease epidemiology of households with similar demographic structures clustered in space we develop} a multi-scale epidemic model consisting of neighbourhoods of households. In our analysis we focus on key parameters which control household size, the importance of transmission within households relative to outside of them, and the degree to which the non-household transmission is localised within neighbourhoods. We construct the household reproduction number $R_*$ over all neighbourhoods and derive the analytic probability of an outbreak occurring from a single infected individual in a specific neighbourhood. We find that reduced localisation of transmission within neighbourhoods reduces $R_*$ when household size differs between neighbourhoods. This effect is amplified by larger differences between household sizes and larger divergence between transmission rates within households and outside of them. However, the impact of neighbourhoods with larger household sizes \boldtext{on an individual's risk of infection is mainly limited to the individuals that reside in those} neighbourhoods.
We consider various surveillance scenarios and show that household size information from the initial infectious cases is often more important than neighbourhood information while household size and neighbourhood localisation influences the sequence of neighbourhoods in which an outbreak is observed.}


\maketitle

\section{Introduction}\label{sec-intro}
The spatial demography of a population influences infectious disease epidemiology in numerous ways. Fundamentally, infection spreads via contact between individuals. Who contacts whom is dependent on the spatial arrangement of individuals in the population. Individuals who share the same school, workplace, or household, for example, have a higher probability of contacting one another than those who do not. The same principle applies to those who share the same social groups.

The spatial separation of individuals with respect to where they live is arguably one of the most significant demographic characteristics to account for. Contact rates are likely to be higher between individuals who live in the same community, neighbourhood or household. Moreover, we may observe spatial clustering of households with similar demographic characteristics, such as household size, density, living conditions and wealth into neighbourhoods or communities. For example, a survey in Ile-Ife, Nigeria reported pockets of high household size associated with areas identified as low income \cite{ibitoye2017spatial} and a report on Ulaanbaatar, Mongolia, conducted by World Bank, found that average household size ranges between $3$ and $5$ in different districts of the city \boldtext{(\cite{kamata2010mongolia}). These observations suggest that a better understanding of the epidemiological implications of multiscale demographic heterogeneities can support improved strategies for the management and prevention of infectious diseases. Here, we focus on districts or neighbourhoods based around clusters of similarly sized households.} We use a multi-scale model which accounts for contact within households, within neighbourhoods and in the wider population to examine how the interplay of transmission patterns at these different scales shapes the epidemiological dynamics. We now briefly review the model paradigms we employ at each scale. 

Neighbourhood structure is captured in a metapopulation framework. The basic principle of a metapopulation model is to partition the total population into subpopulations based on their spatial separation. These types of models were first introduced by Levins and subsequently developed by Hanski to study ecological phenomena \boldtext{(\cite{levins1969some,hanski1998metapopulation}). They have been used in many studies of epidemiological dynamics including recently to model infection spread through a population of villages containing households (\cite{britton2011inference}), to model the emergence of a novel pathogen spreading from a village to a nearby city (\cite{kubiak2010insights}), and the development of multi-scale hierarchical epidemic models (\cite{watts2005multiscale}).} 

\boldtext{The framework we present here is similar to the special case presented in example 1 of \cite{britton2011inference}, a study that focuses on statistical inference. Our work moves beyond that example by allowing for different household sizes in each neighbourhood, introducing heterogeneity in the contact rates between neighbourhoods, and examining the implications of the population structure for disease surveillance and control.} 

\boldtext{\cite{kubiak2010insights} use a metapopulation model to study the emergence of an infectious disease in a rural village and its subsequent spread to a nearby city via commuter travel. They find that spatial heterogeneity only has a limited effect on the probability of emergence and outbreak size. The work we present here complements this study by accounting for the finer scale structure associated with transmission within households in addition to the structure at the neighbourhood or village scale.}

\boldtext{\cite{watts2005multiscale} consider a hierarchical stochastic epidemic model which incorporates mixing on multiple scales. Each scale represents a different community grouping, increasing in size, such as neighbourhoods, cities and regions of a country. Individuals make contact with one another at their local context only. Our model has some similarities with \cite{watts2005multiscale} in that it considers spatial structure on several scales. However, the structures in our model incorporate concepts of residency; an individual's household remains fixed, we distinguish between contact within the household and outside of it, within an individual's own neighbourhood and beyond it.} 

Household structure is typically captured in stochastic model frameworks based on small groups of cohabiting individuals. Household models are ideal for distinguishing between transmission within a small well-defined group such as a family and transmission in the wider community. In particular, household structure is important when transmission is strong within the household but weak in the wider community. This is often the case as we usually expect contacts within a household unit to be closer, more frequent and longer. Household models date back to the 1970s \boldtext{(\cite{bartoszynski1972certain})} and remain an active field of research. They can be formulated in a number of ways. Early work focused on stochastic processes, acknowledging the small number of individuals in the household. Extensive work has been carried out in developing probabilistic frameworks for household models \boldtext{(\cite{ball1997epidemics,ball2001stochastic,ball2002general})}, which model the infectious contacts occurring between individuals from different groups as points on a Poisson process.

In these frameworks epidemic thresholds are typically derived using branching process theory. A range of reproduction numbers have been explored, of which the household reproduction number $R_*$, is usually the most intuitive and tractable \boldtext{(\cite{pellis2012reproduction,ball2016reproduction})}.  
$R_*$ is the expected number of households with at least one infected individual resulting from a single infected household in an otherwise susceptible population. \boldtext{Other important quantities include the initial growth rate of the epidemic $r$, the final epidemic 
size and the household offspring distribution. \boldtext{This is the distribution of the number of secondary infected households resulting from one infectious household (\cite{ross2010calculation})}.}

Stochastic compartment models can also be formulated using systems of ordinary differential equations (ODEs), often referred to as master or forward Kolmogorov equations. \boldtext{Here, the state variables define the probability that a household is in given epidemiological state and the ODEs define the transitions between the possible states (\cite{house2008deterministic,ross2010calculation})}. 
\boldtext{\cite{ross2010calculation} present fundamental theory based on describing the chain of infection within the household as a Markov process. They show that large households can act as amplifiers of infection. \cite{black2013epidemiological} consider a similar model to investigate household-based antiviral treatment during an influenza pandemic. However, they assume a heterogeneous distribution of households and consider several different scenarios for the delay before the household antiviral treatment has an impact. \cite{adams2016household} uses the same framework \boldtext{as \cite{ross2010calculation, black2013epidemiological}} to investigate Ebola epidemiology. In agreement with \cite{ross2010calculation}, they find that intense within-household transmission makes communities composed of large households uniquely vulnerable to Ebola outbreaks.}

\boldtext{Other studies have included the effects of demographic change. Frameworks have been developed that account for evolving age and household structure both using agent-based approaches (\cite{geard2015effects}) and systems of ODEs (\cite{hilton2019incorporating}). \cite{hilton2019incorporating} find that age-structured assortative contact focuses transmission in younger age-groups, while household contact focuses transmission in family groups. These two structures act synergistically to concentrate infection in large households with a majority of young people.}

We employ the ODE approach to describe the epidemiological dynamics of a system composed of neighbourhoods of households. The underlying structure of our model has been analysed from a rigorous probabilistic perspective by \boldtext{(\cite{ball2001stochastic})}. That study is based on a very general stochastic multitype household model in which individuals are differentiated into classes, and grouped into households. Here we cast those population classes as neighbourhoods and use the ODE framework to bridge to a more applied context with population structures inspired by those observed in cities in Mongolia and Nigeria \boldtext{(\cite{kamata2010mongolia,ibitoye2017spatial})}.

\boldtext{In summary, there is a solid history of using metapopulation and household models to explore the role of demographic structures in epidemiological dynamics. However, gaps remain and more work is required to reach a comprehensive understanding. In particular, studies have examined neighbourhood scale structure but either omitted household structure (\cite{kubiak2010insights}),  lacked the flexibility to vary household sizes between neighbourhoods (\cite{britton2011inference}), or used multiscale configurations that do not account for residency (\cite{watts2005multiscale}).}

\boldtext{Here, we present a multiscale model based on a master equation formulation and analyse it using branching process approximations and Markov processes theory, in addition to numerical methods such as the Gillespie SSA. We explore how the household size in each neighbourhood, the relative importance of transmission within households versus outside of them, and the localisation of transmission outside households affect (i) the risk of an epidemic in the population (ii) an individual's risk of infection (iii) where outbreaks are first observed.}

\section{Model Formulation}\label{sec-model}
\begin{figure}[]
\centering
\includegraphics[width=0.8\textwidth]{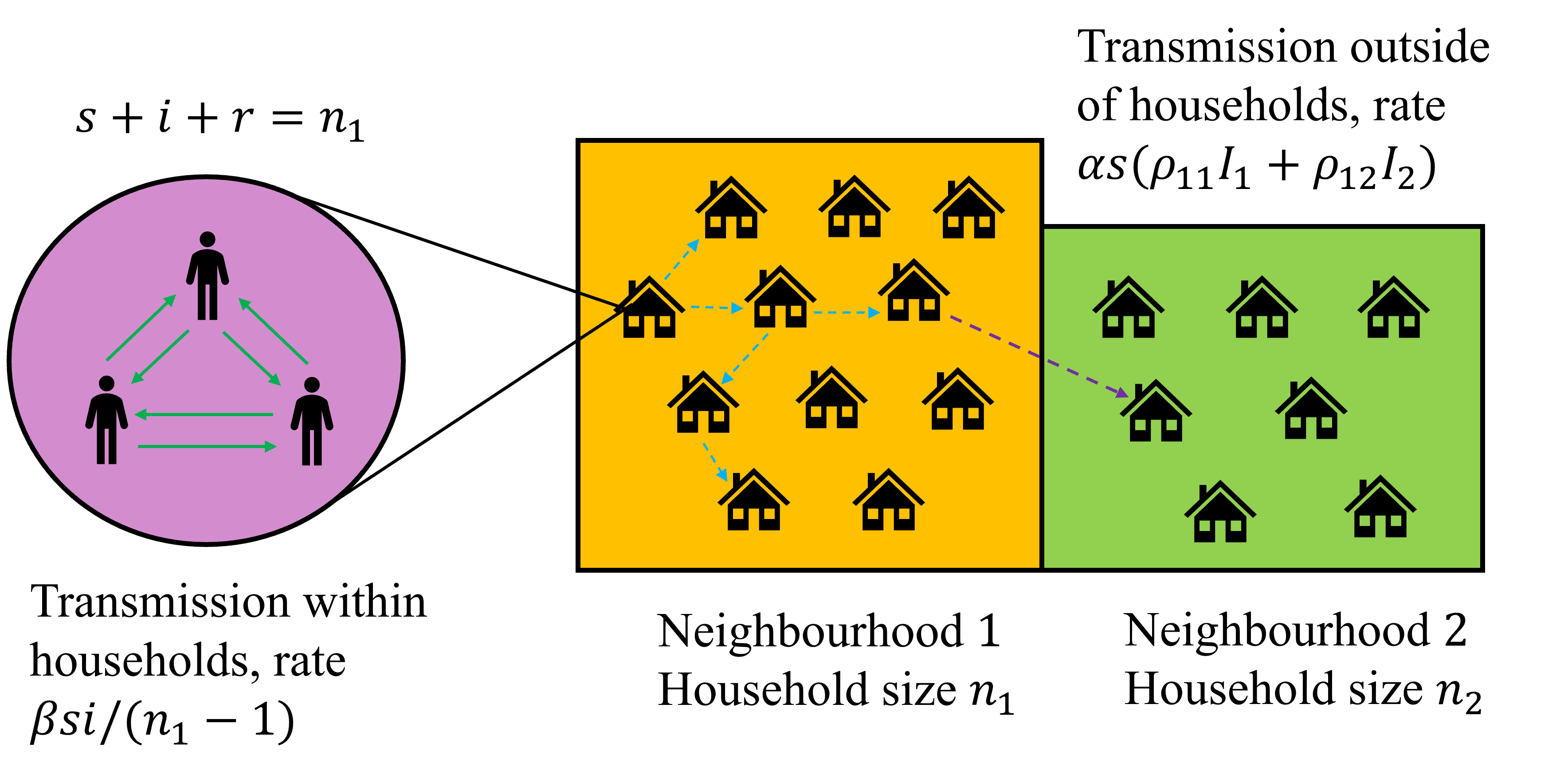}
\caption{Schematic of the two neighbourhood model. There are two levels of connectivity: contacts within households and contacts outside of households either with individuals from an individual's own neighbourhood or from another neighbourhood.}\label{fig-model}
\end{figure}

We consider a model composed of $K$ neighbourhoods, in each of which the population is organised into households (see Figure \ref{fig-model}). In this study, the total population size in neighbourhood $i$ is $N_i=N$, for $i = 1,\ldots K$ \boldtext{so that each neighbourhood has equal population size}. The household size in neighbourhood $i$ is denoted $n_i$. This parameter may vary between neighbourhoods, \boldtext{but we restrict each household in a given neighbourhood to be the same size.}
Hence, the total population size is the sum of the neighbourhood populations $\sum_i N_i= \sum_i n_ih_i$, where $h_i$ denotes the number of households in neighbourhood $i$. In our framework, contact occurs at two scales. All contact rates are frequency dependent. Individuals make contact with members of their household at rate $\beta$. This parameter takes the same value in all neighbourhoods. Individuals make contact outside of their household at rate $\alpha$. A proportion $\rho_{ii}$ of these contacts are with individuals from their own neighbourhood $i$ and a proportion $\rho_{ij}$ are with individuals from neighbourhood $j$.

\subsection{Neighbourhood localisation}
We model the localisation of contact within neighbourhoods by expressing $\rho_{ij}$, the proportion of contacts an individual from neighbourhood $i$ has with individuals from neighbourhood $j$, in terms of the proportion of contacts which are `reserved' for individuals from their `home' neighbourhood, $r_i$ \boldtext{(\cite{jacquez1988modeling})}.  
Contacts outside of the home neighbourhood are then allocated proportionally to the neighbourhood sizes such that 
\begin{align*}
\rho_{ii} = r_i + (1-r_i)\frac{\alpha(1-r_i)N_i}{\sum_k \alpha(1-r_k)N_k} = r_i + (1-r_i)\frac{(1-r_i)N_i}{\sum_k(1-r_k)N_k},
\end{align*}
and 
\begin{align*}
\rho_{ij} = (1-r_i)\frac{(1-r_j)N_j}{\sum_k(1-r_k)N_k},
\end{align*}
where the sums are over $k = 1,\ldots K$. Proportional mixing between the neighbourhoods corresponds to $r_i=0$ for all neighbourhoods. In this case no contacts are reserved exclusively for an individual's home neighbourhood.
If all neighbourhoods are of equal size, $\rho_{ii}=\rho_{ij}=1/K$, where $K$ is the number of neighbourhoods. Complete isolation of neighbourhoods corresponds to $r_i=1$. In this case all contacts are reserved for an individual's home neighbourhood. Varying $r_i$ from $0$ to $1$ increases the neighbourhood contact localisation from proportional mixing to complete isolation.

\subsection{Households}
The household populations are structured according to the deterministic framework of \boldtext{\cite{house2008deterministic,adams2016household}}. Here, each individual in the household is categorised by their disease state: susceptible $(s)$, infectious $(i)$ or recovered $(r)$. Demographic processes are omitted. The state or configuration of a household $(s,i,r)$ is then defined by the number of individuals in the household in each disease state. The set of all possible household states in neighbourhood $j$ is $\mathcal{S}_j=\{(s,i,r)\vert 0\leq s,i,r \leq n_j \text{\;and\;}s+i+r= n_j\}$. We define $\mathcal{C}$ to be the set of all transient household states; \boldtext{those states that have at least one infectious individual.}  

A household transitions from one state to another when either (1) a household member becomes infected or (2) an infected household member recovers. Infectious individuals recover at rate $\gamma$.
Transmission of infection can occur through either contact within the household or contact outside of the household. Hence, the force of infection acting on a susceptible individual from a household in neighbourhood $i$ is $\alpha(\rho_{ii}I_i + \sum_j \rho_{ij}I_j)+\frac{\beta i}{s+i+r-1}$ where $I_j$ is the proportion of the population in neighbourhood $j$ that is infected. We explain the construction of the transmission rate in Section \ref{sec:house-model}. We model the transmission process within the household as a continuous-time Markov process (CTMP) \boldtext{(\cite{ross2010calculation})}. We let $q(k,l)$ be the rate of transition from household state $k$ to state $l$ if $k\neq l$ and $q(k,k)=-\sum_{k\neq l}q(k,l)$. These transition rates form the elements of the Markovian transition matrix $Q$. The rates of all possible transitions between states are shown in Table \ref{tab1}.

\begin{table}[h]
\begin{center}
\begin{minipage}{280pt}
\caption{Transition rates from household state $k$ to state $l$ (for a household in neighbourhood $1$). The transition due to infection can occur by transmission within the household, or outside of it.}\label{tab1}
\begin{tabular}{@{}llll@{}}
\toprule
State $k$ & State $l$  & Transition & Rate $q(k,l)$\\
\midrule
$(s,i,r)$    & $(s-1,i+1,r)$  & infection  & $\alpha s(\rho_{11}I_1+\sum_j\rho_{1j}I_j)+\frac{\beta si}{s+i+r-1}$  \\
$(s,i,r)$     & $(s,i-1,r+1)$   & recovery & $\gamma i$  \\
\botrule
\end{tabular}
\end{minipage}
\end{center}
\end{table}

\subsection{Households of size 1}
Here we generate some insight into the neighbourhood structure by considering a model with two neighbourhoods composed of households of size $1$. So $n_1=n_2=1$. We denote the total proportions of susceptible, infectious and recovered individuals in neighbourhood $i$ by $S_i$, $I_i$ and $R_i$ respectively. Then
\begin{align}\label{eq:metapop}
\begin{split}
\frac{dS_1}{dt}&=-\alpha \left( \rho_{11} I_1+\rho_{12} I_2\right)S_1, \\
\frac{dS_2}{dt}&=-\alpha\left( \rho_{22} I_2+\rho_{21} I_1\right)S_2, \\
\frac{dI_1}{dt}&=\alpha\left(\rho_{11} I_1+\rho_{12} I_2\right)S_1 - \gamma I_1, \\
\frac{dI_2}{dt}&=\alpha\left(\rho_{22} I_2+ \rho_{21} I_1\right)S_2 - \gamma I_2, \\
\frac{dR_1}{dt}&=\gamma I_1, \\
\frac{dR_2}{dt}&=\gamma I_2.
\end{split}
\end{align}
The total population size remains constant and $S_i + I_i +R_i =1$ for $i=1,2$. The total number of contacts per unit time individuals from neighbourhood $i$ make with individuals from neighbourhood $j \neq i$ must be equal to the number of contacts individuals from neighbourhood $j$ make with individuals from neighbourhood $i$. Therefore $\rho_{12}N_1=\rho_{21}N_2$, where $N_1$ and $N_2$ are the respective population sizes of neighbourhoods $1$ and $2$. We assume $\alpha$ is the same for both neighbourhoods. 

The basic reproduction number for this model, found using next generation matrix methods \boldtext{(\cite{diekmann2000mathematical})}, is $R_0=\frac{\alpha}{2\gamma}\left[(\rho_{11}+\rho_{22})+\sqrt{(\rho_{11}+\rho_{22})^2-4(\rho_{11}\rho_{22} - \rho_{12}\rho_{21})}\right]$. Clearly, under proportional mixing and when neighbourhoods are of equal size, $\rho_{11}=\rho_{12}=\rho_{21}=\rho_{22}=1/2$ and the reproduction number reduces to $R_0=\alpha/\gamma$. However, when mixing is not proportional the neighbourhood structuring influences $R_0$.

Therefore, in the absence of household structure, neighbourhood localisation of contact can influence the epidemic risk and the expected numbers of infections in the early generations of an epidemic.

\subsection{Households of size $n>1$}\label{sec:house-model}
For two neighbourhoods  $j=1,2$ with households of fixed size $n_j$, we can describe the epidemiological dynamics in the population as a whole using the single household approximation \boldtext{(\cite{house2008deterministic,holmes2022approximating}) given by the system of ODEs:}
\begin{align}\label{eq:coupledODE}
\begin{split}
 \dot{H}^{j}_{sir}&=\alpha(\rho_{jj} I_j+\rho_{jk} I_{k})\left[-sH^j_{sir}+(s+1)H^j_{(s+1)(i-1)r} \right]\\
    &+\beta\left[-s\frac{i}{n_j-1}H^j_{sir}+(s+1)\frac{i-1}{n_j-1}H^j_{(s+1)(i-1)r}\right]\\
    &+\gamma \left[-iH^j_{sir} + (i+1)H^j_{s(i+1)(r-1)}\right].
    \end{split}
\end{align}
Here, $H^j_{sir}$ denotes the probability that a household is in neighbourhood $j$ in state \boldtext{$(s,i,r)\; \in \; \mathcal{S}_j$} and $I_j=\frac{\sum_{\mathcal{S}_j}i\cdot H_{sir}^j}{n_j}$ is the proportion of individuals in neighbourhood $j$ that are infected at time $t$. Other household states and proportions are denoted analogously.
\boldtext{In system \eqref{eq:coupledODE} we adopt the convention that the probability $H^j_{sir}$ associated with any state $(s,i,r)$ is set to zero if the state is not contained in $\mathcal{S}_j$, the state space of neighbourhood $j$, for $j=1,2$.} For example, for states with either $s,i,r<0$, the proportion of households in these states is automatically zero. The terms in the square brackets correspond to transmission outside of the household, transmission within the household and recovery respectively.

The terms for transmission outside of the household involve several variables. Each of the $s$ susceptibles in a given household contacts individuals from the wider community at rate $\alpha$. For a household in neighbourhood $j$, transitions between states $(s,i,r)$ and $(s-1,i+1,r)$ due to transmission outside of the household occur at rate $\alpha s(\rho_{jj} I_j+\rho_{jk} I_{k}), k \ne j$. Within a household, each of the $s$ susceptible individuals contacts other individuals in the household at rate $\beta$. A proportion $i/(n_j-1)$ of these contacts are with infectious individuals. Hence the within-household transmission rate for neighbourhood $j$ is $\beta si/(n_j-1)$.

\subsection{Calculating $R_*$}\label{subsec:Rstarcalc}
The household reproduction number $R_*$ is defined as the expected number of households infected by a single infectious household, in an otherwise susceptible population. $R_*$ is usually found by approximating the early stages of the population-wide epidemic by a household-level branching process \boldtext{(\cite{pellis2012reproduction,ball1997epidemics})}. It is possible to derive closed form expressions for $R_*$ for a single neighbourhood of households of size $2$ or $3$; see Supplementary \boldtext{information} (Section $1$). However, for large household sizes and multiple neighbourhoods, expressions for $R_*$ become more complex. We therefore calculate $R_*$ for general household sizes $n_j$ using the transition matrix $Q$ for the within-household model. Full details of this construction are given in \cite{ross2010calculation}; however, we briefly explain the concept below.

\boldtext{In the initial exponential growth phase of the epidemic the proportion of households of each size among all newly-infected households is fixed in the `stable ratio'. So} to obtain the expected number of infected households originating from a single infectious household, we average over all neighbourhoods weighted by this stable ratio in the exponential growth phase of the epidemic. For two neighbourhoods this corresponds to taking the spectral radius of the next generation matrix $K$ \boldtext{(\cite{diekmann2000mathematical})},
\begin{align}
   K= \begin{pmatrix}
    R^{11}_* & R^{12}_* \\
    R^{21}_* & R^{22}_*
    \end{pmatrix},
\end{align}
where $R^{ij}_*$ is the expected number of infected households in neighbourhood $j$ arising from a single primary infected household in neighbourhood $i$ in an otherwise susceptible population.

$R^{11}_*$, for example, can be written as $\frac{\alpha \rho_{11}}{\gamma}H^{1,\infty}_j$. Here, $H^{1,\infty}_j$ is the expected household epidemic size (of a household in neighbourhood 1), where initially there is one infected individual in the household (state $j=n_1-1,1,0$). Following \cite{ross2010calculation,adams2016household}, we can write $H^{1,\infty}_j$ as $\gamma e^1_j$, where $e^1_j$ is the expectation of the path-integral $\Gamma=\int_{0}^{\infty}f(X(t))dt$ conditional on the Markov process starting in state $j$. Here $X(t)$ is the CTMP that describes the infection process within the given household and $f(j)$ maps the household state $j$ to the number of infectious individuals in that state $j$. So $f(j)=i$ when $j$ is state $(s,i,r)$.
Each time a new individual becomes infected within the household, the \boldtext{path-integral $\Gamma$} increases by $1/\gamma$, the expected infectious period. Consequently, the expected household epidemic size for a household in neighbourhood $1$, starting in state $j=n_1-1,1,0$, is $H^{1,\infty}_j=\gamma e^1_j$. The expected number of infections an infected individual in neighbourhood $1$ produces outside of their own household but within their home neighbourhood is $\alpha \rho_{11} / \gamma$. We assume that each of these infections involves someone from a different household. Following \cite{ross2010calculation}, we find $e^1_j$ by solving the following system of linear equations
\begin{align}\label{eq:lin1}
    \sum_{k\in \mathcal{C}_1}q^1(j,k)e^1_k+f(j)=0, \;\;\; j \in \mathcal{C}_1.
\end{align}
In equations \eqref{eq:lin1}, $\mathcal{C}_1$ denotes the set of all transient states for a household in neighbourhood $1$; $q^1(j,k)$ is the rate of transition from household state $j$ to $k$ in neighbourhood $1$ and $f(j)$ is defined as above.

In conclusion, the expected number of infected households in neighbourhood $1$ arising from one infected household in neighbourhood $1$ is 
\begin{align*}
    R^{11}_* = \frac{\alpha \rho_{11}}{\gamma}H^{1,\infty}_j = \frac{\alpha \rho_{11}}{\gamma} \gamma e_j^1 = \alpha \rho_{11} e_j^1.
\end{align*}
Similar considerations give
\begin{align*}
R^{12}_*&=e^1_j\alpha \rho_{12}=e^1_j\alpha (1-\rho_{11}),\\
R^{21}_*&=e^2_j\alpha \rho_{21}=e^2_j\alpha (1-\rho_{22}),\\
R^{22}_*&=e^2_j\alpha \rho_{22}.
\end{align*}

\subsection{Probability of an outbreak}\label{subsec:prob_outbreak}
\subsubsection{A single neighbourhood}
Consider the branching process for the single neighbourhood model. Under a branching process framework there are two possible outcomes: the outbreak becomes extinct with probability $1$ or there is a positive probability that the number of infections grows without bound \boldtext{(\cite{diekmann2000mathematical,athreya1972branchingprocesses,lloyd2007stochasticity})}.
We define the probability of a major outbreak as the probability that the branching process does not become extinct (for a given initial condition on the number of infected individuals in the neighbourhood(s)).
\boldtext{For a single neighbourhood population, he branching process originating from a single infectious individual dies out with probability $p$, where $p$} is the minimal non-negative solution of the equation $G(p)=p$. So $1-p$ is the probability of a major outbreak. Here $G(p)=\sum_{k=0}^{\infty}p^k g(k)$ is the offspring probability generating function and $g(k)$ is the probability that a primary infectious household produces $k$ secondary infectious households.

These probabilities correspond to the offspring distribution for the given household of size $n$. Closed form expressions for the offspring distribution are relatively easy to obtain for households of size $1$ and $2$; see Supplementary information (Section $2$). For larger household sizes we find the offspring distribution numerically. This involves solving a system of linear equations \boldtext{(\cite{pollett2002path,ross2010calculation})} in order to obtain the Laplace Stieltjes transform (LST) $y_i(s)=\mathbbm{E}[\exp{(-s \alpha \Gamma)}\vert X(0)=i]$ of the path-integral $\Gamma$ for $i$ in the set of all transient household states $\mathcal{C}$. This is the same path-integral as in Subsection \ref{subsec:Rstarcalc}; recall that $X(t)$ is the CTMP that describes the infection process within the household and $i$ is the initial state of the household.
The system of linear equations is
\begin{align}\label{eq-ross2}
    \sum_{j\in \mathcal{S}} q(i,j)y_j(s)=s\alpha f(i) y_i(s),\;\;\; i \in \mathcal{C}.
\end{align}
As previously, $q(i,j)$ represents the rate of transition from household state $i$ to $j$ for a single neighbourhood model. The set of all household states is $\mathcal{S}$ and $\mathcal{C}$ denotes the set of all the transient household states.
From this, the required offspring distribution can be constructed recursively by noting that $g(k)=(-1)^{k}\frac{y_1^k(1)}{k!}$ and we can obtain the $k\textsuperscript{th}$ derivative of $y_1(s)$ with respect to $s$ by differentiating system \eqref{eq-ross2} \boldtext{(\cite{ross2010calculation})}.

Under the branching process framework, we assume an infinite population of households. In consequence, each new infection outside of a household occurs in a fully susceptible household that has not been previously infected. This is a reasonable approximation in the early stages of an outbreak \boldtext{(\cite{ball1997epidemics,ball2008control})}.

\subsubsection{Two neighbourhoods}\label{sec-prob-neigh}
For the two neighbourhood model, we construct a multi-type branching process of neighbourhood $1$ and neighbourhood $2$ households. Let $X_1(t)$ be the CTMP describing the infection process occurring within a household in neighbourhood $1$. We define $\Lambda_{11}$ as the total force of infection on households in neighbourhood $1$ over the course of the neighbourhood $1$ household epidemic:
\begin{align*}
    \Lambda_{11}=\int^{\infty}_0f_1(X_{1}(t))dt.
\end{align*}
$f_1(X_{1}(t))$ is the rate at which a household in neighbourhood $1$ infects other households in neighbourhood $1$ (conditional on the initial state of the Markov process). $\Lambda_{12}$ is defined analogously,
\begin{align*}
    \Lambda_{12} = \int_0^{\infty} f_2(X_1(t)) dt.
\end{align*}
This is the total force of infection experienced by households in neighbourhood $2$ over the course of the neighbourhood $1$ household epidemic. The joint LST of $(\Lambda_{11},\Lambda_{12})$, conditioned on the initial household state, $IC_1=(s=n_1-1,i=1,r=0)$, is the total force of infection outside of the household over the course of the (neighbourhood $1$) household epidemic \boldtext{(\cite{ross2015contact})}. This is
\begin{align}\label{eq:ysyst}
    y_{j}^1(s_1,s_2)=\mathbbm{E}\left[\exp{(-s_1\Lambda_{11}-s_2\Lambda_{12})} \vert X_1(0)=j\right], 
\end{align}
where $j=IC_1$ is the initial household state.

Adapting the results of the path-integral methods used in \cite{pollett2002path,ross2010calculation,ross2015contact} it follows that,
\begin{align}\label{eq:lin-sys2}
    \sum_{j \in \mathcal{S}_1} q^1(i,j)y_j^1(s_1,s_2)&=(f_1(i)s_1 +f_2(i)s_2)y_i^1(s_1,s_2)\\
    &=\alpha (\rho_{11} s_1 +\rho_{12} s_2)I(X_1=i)y_i^1(s_1,s_2), \;\;\; \forall i \in \mathcal{C}_1.
\end{align}
$q^1(i,j)$, $\mathcal{S}_1$ and $\mathcal{C}_1$ are defined as in Subsection \ref{subsec:Rstarcalc}. The function $I(X_1=i)$ is used to denote the number of infectious individuals in the household when the CTMP is in state $i$.

Let the probability that a chain of infection starting from a single individual of type $m$ becomes extinct be $p_m$. Then in a multi-type branching process of two types, $p_m$ is the minimal non-negative solution to the system of equations
\begin{align*}
    p_m = G_m(p_1,p_2),\;\;\; m=1,2,
\end{align*}
where $G_m$ is the offspring probability generating function of an individual of $m\textsuperscript{th}$ type. For the neighbourhood $1$ type, the generating function is
\begin{align}\label{eq:pgf}
    G_1(p_1,p_2)=\sum_{k_1=0}^{\infty}\sum_{k_2=0}^{\infty}p_1^{k_1}p_2^{k_2}g(k_1,k_2),
\end{align}
where $g(k_1,k_2)$ is the probability of $k_1$ and $k_2$ newly infected households in neighbourhoods $1$ and $2$ respectively arising from the household epidemic in a single infected household in neighbourhood $1$. This probability can be written as 
\begin{align}\label{eq:infection-force-pdf}
    g(k_1,k_2)=\int_0^{\infty}\int_0^{\infty}\eta_1(\lambda_1,\lambda_2)\frac{\exp{(-\lambda_1)}\lambda_1^{k_1}}{k_1!}\frac{\exp{(-\lambda_2)}\lambda_2^{k_2}}{k_2!}d\lambda_1d\lambda_2,
\end{align}
where $\eta_1(\lambda_1,\lambda_2)$ is the joint probability density function of the total force of infection (from transmission outside of the households) acting on households in each neighbourhood over the course of the epidemic in the initial infected household in neighbourhood $1$. 

If we substitute \eqref{eq:infection-force-pdf} into \eqref{eq:pgf}, we are able to swap the integration and summations in order to achieve
\begin{align}\label{eq:}
    G_1(p_1,p_2)=\int_0^{\infty}\int_0^{\infty}\eta_1(\lambda_1,\lambda_2)\exp{(-\lambda_1(1-p_1))}\exp{(-\lambda_2(1-p_2))}d\lambda_1 d\lambda_2;
\end{align}
remembering also the power series expansion for the exponential function. Therefore, the offspring generating function is equal to $y^1_{IC}(1-p_1,1-p_2)$, where $IC$ \boldtext{denotes} the initial condition. So the extinction probability when starting with an initial infection in neighbourhood $1$, $p_1$, is given by the minimal non-negative solution $(p_1,p_2)$ to the system of equations
\begin{align}\label{eq:BP}
    p_1=y^1_{IC_1}(1-p_1,1-p_2)\;\;\ \text{and} \;\;\; p_2=y^2_{IC_2}(1-p_1,1-p_2),
\end{align}
where $IC_1=(s=n_1-1,i=1,r=0)$ and $IC_2=(s=n_2-1,i=1,r=0)$ for neighbourhoods $1$ and $2$ respectively and $y^2$ is defined analogously to $y^1$. The set of equations \eqref{eq:BP} can be solved numerically and the joint LST is obtained by solving the linear system of equations \eqref{eq:lin-sys2} for $(s_1,s_2)=(1-p_1,1-p_2)$.

This methodology can be generalised to $k$ neighbourhoods, and to different household types within neighbourhoods \boldtext{(\cite{ross2015contact})}.

\subsection{Parameterisation}
In the results presented in Section \ref{sec-results} we use parameter values from the ranges detailed in Table \ref{tab2}. These values are consistent with an acute respiratory infection such as influenza. The contact rates were determined by first fixing $R_*$, when all household sizes are $2$, to $2.4$ (unless stated otherwise). \boldtext{This value is in line with estimates for $R_*$ in \cite{black2013epidemiological} and references therein.} We do this with the aim of creating a `level playing field' for comparison of results. Next we set the parameter $\nu$. This is the ratio of within household contacts to outside of household contacts, such that $\beta=\nu \alpha$. Larger values of $\nu$ correspond to a greater emphasis on the within-household transmission. Finally, we find the required $\beta$, and hence $\alpha$, values.

\begin{table}[h]
\begin{center}
\begin{minipage}{280pt}
\caption{Parameter values used for all numerical results, unless otherwise stated. These values are consistent with an acute respiratory infection such as influenza \boldtext{(\cite{black2013epidemiological})}. \boldtext{The lower section of the table shows quantities derived from parameters in the upper section.} All rates per day.}\label{tab2}%
\begin{tabular}{@{}lll@{}}
\toprule
Parameter & Meaning  & Values used \\
\midrule
$n_i$ & household size in neighbourhood $i$  & $2$--$6$  \\
$h_i$ & no. households in neighbourhood $i$, $\sum_{i=1}^2n_i h_i=5040$ & $840$--$2520$ \\
$r_i$ & propn. external contacts reserved for own neigh. & $0$--$1$ \\
$\nu$ & ratio of within to outside of household contacts & $1$--$5$ \\
$\gamma$ & recovery rate & $0.2$ \\
$R_*$ & household reproduction number & $2.4$ \\
\midrule
$\alpha$ & transmissible contact rate outside of households & $0.25$--$0.3$ \\
$\beta$ & transmissible contact rate within households & $0.3$--$1.29$ \\
\botrule
\end{tabular}
\end{minipage}
\end{center}
\end{table}

\subsection{Model analysis}
\boldtext{Our model analysis focuses on understanding the impact that three key parameters have on the epidemiological dynamics: household size $n_i$, neighbourhood localisation $r_i$ and the ratio of within household contacts to outside of household contacts $\nu$. In particular, we investigate how these quantities influence the household reproduction number $R_*$, individual infection risk, the probability of an outbreak and the probability that an outbreak is first observed in a given neighbourhood. We also consider the implications of our model for surveillance and control strategies, particularly the importance of accounting for the neighbourhood or household demography of the first cases that are detected.} 

\boldtext{In Subsections \ref{subsec:neigh_loc_Rstar} and \ref{subsec-surv} we use the analytic approaches from Section \ref{sec-prob-neigh} to calculate outbreak probabilities. In Subsection \ref{subsec:neigh_loc_ind_risk} we use the single household approximation ODE model (system \eqref{eq:coupledODE}) to determine the individual infection risk for residents of each neighbourhood. In Subsection \ref{subsec:outprob_surv} we compare outbreak probabilities calculated under various scenarios using analytic and numerical (SSA) methods (\cite{gillespie1977exact}). Finally, in Section \ref{sec:6neigh} we use a numerical methodology to explore the impact of more complex neighbourhood structures on the infectious disease epidemiology. In the majority of the paper we consider models with two neighbourhoods, but in Section \ref{sec:6neigh} we consider six neighbourhoods.}

\subsubsection{Individual infection risk}
\boldtext{We define the individual infection risk as the proportion of individuals in the neighbourhood, or in the population as a whole, that have been infected after $100$ days. For our parameters, after $100$ days the outbreak has ended, prevalence is very low and the effective reproduction number is less than $1$. We chose this metric instead of the more commonly used final epidemic size because it has an intuitive interpretation and allows for a clear comparison between residents of different neighbourhoods. We calculate the individual infection risk by solving the master equation model \eqref{eq:coupledODE} over the specified time period and finding the final proportion of individual residents in each neighbourhood that are in the recovered state $r$.}

\subsubsection{Outbreak probability}
\boldtext{We calculate outbreak probabilities using the analytic approaches described in Section \ref{sec-prob-neigh} and numerically by using a Gillespie SSA (\cite{gillespie1977exact}) to simulate the infection process, as described by the coupled differential equations in system \eqref{eq:coupledODE}, for a large number of independent trials in which a single infected individual is introduced into a naive population. The simulated populations are finite and there are often no households in a particular state. So, in order to control the computational complexity, we maintained a dynamic set of household states. We used a dictionary to define the set of all household states present in each neighbourhood at time $t$. Propensity functions were defined for each household state and infection or recovery event. As events occurred, the household dictionary was updated, removing and adding new household states as necessary.} 

\subsubsection{More complex neighbourhood structures}\label{subsec:methods_6neigh}
\begin{figure}
    \centering
    \includegraphics[width=0.5\linewidth]{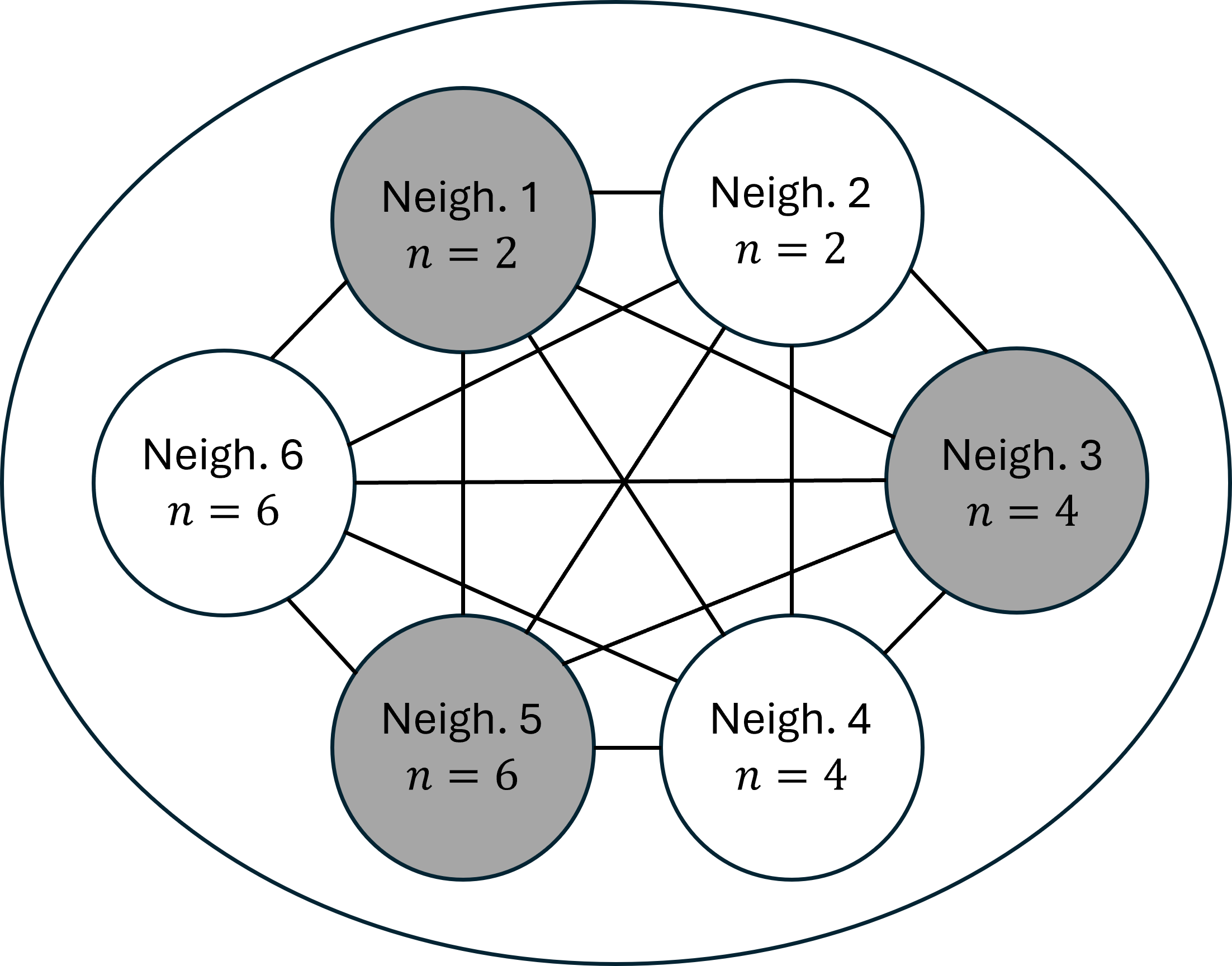}
    \caption{\boldtext{Schematic of the $6$ neighbourhood model. The bigger circle represents the total population and the smaller circles denote each neighbourhood. The neighbourhoods are numbered as in Table \ref{tab-neighs} with neighbourhood household sizes detailed in each. The lines denote the connectivity between all the neighbourhoods. The shading of the neighbourhoods distinguish between the strengths of the localisation of their contacts with all other neighbourhoods. The shaded neighbourhoods represent weaker localisation (stronger connectivity) and non-shaded neighbourhoods hold stronger localisation (weaker connectivity).}}
    \label{fig:6neigh}
\end{figure}

\boldtext{In order to explore how our key parameters, such as household size and the degree to which non-household transmission is localised within neighbourhoods, influence the epidemiological dynamics under more complex neighbourhood structures, we consider a model with $6$ neighbourhoods. Our Gillespie SSA code readily accommodates any number of neighbourhoods. We set up a model composed of two neighbourhoods with households of size $2$, two neighbourhoods with households of size $4$ and two neighbourhoods with households of size $6$. The total number of individuals in each neighbourhood is fixed at $N$. For the principle set of trials, the parameters that control localisation of contact outside of the household in each of these neighbourhoods are randomly assigned such that, in each pair of neighbourhoods, one has weakly localised contacts ($r\sim U[0,0.1]$) and the other has strongly localised contacts ($r\sim U[0.4,0.5]$). See Table \ref{tab-neighs} for the summary of the principle neighbourhood set up \boldtext{and Figure \ref{fig:6neigh} for a schematic representation of the model}. Structuring the model in this way aims to strike a balance between generality and computational tractability. In a second set of trials, the localisation parameters for all neighbourhoods are drawn from the same distribution $r\sim U[0,0.7]$. Within these frameworks we investigate the impact that demographic structure has on the probability that an outbreak is first observed in a given neighbourhood, and the sequence of neighbourhoods in which the outbreak becomes apparent.}

\begin{table}[h]
\begin{center}
\begin{minipage}{330pt}
\caption{Six neighbourhood model structure for Figures \ref{fig-obs-risk}, \ref{fig-all-seq} and \ref{fig-high-seq}. $U[a,b]$ denotes that the neighbourhood localisation $r_i$ is uniformly distributed over the interval $[a,b]$.}\label{tab-neighs}%
\begin{tabular}{|l|llllll|}
\hline
Neighbourhood & $1$ & $2$ & $3$ & $4$ & $5$ & $6$ \\
\hline
Household size & $2$ & $2$ & $4$ & $4$ & $6$ & $6$ \\
$r_i$ & $U[0,0.1]$ & $U[0.4,0.5]$ & $U[0,0.1]$ & $U[0.4,0.5]$ & $U[0,0.1]$ & $U[0.4,0.5]$\\
\hline
\end{tabular}
\end{minipage}
\end{center}
\end{table}

\begin{figure}
\centering
\includegraphics[width=0.6\textwidth]{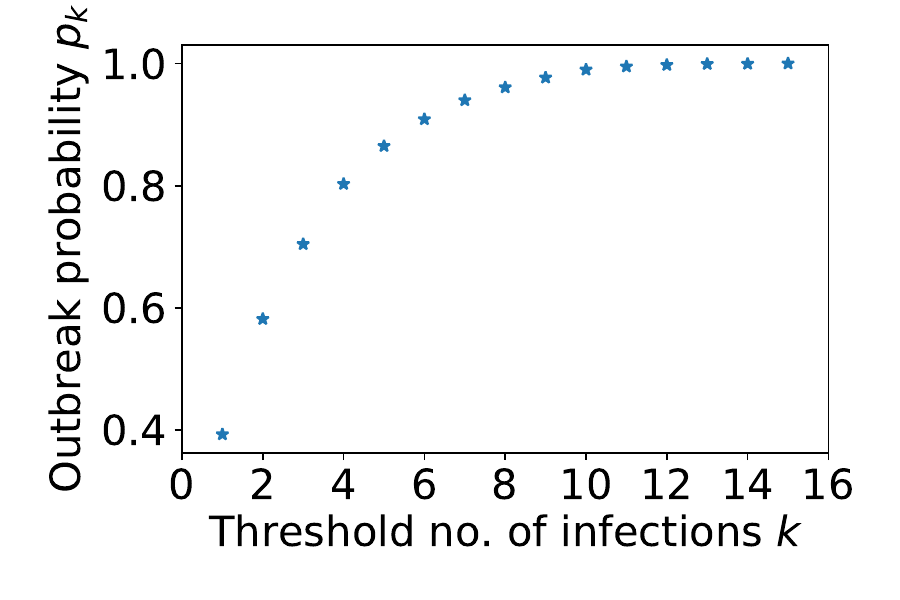}
\caption{The probability $p_k$ of a large outbreak occurring after $k$ household infections have been observed. The Gillespie SSA was used to produce $10,000$ realisations of the model with a single neighbourhood composed of households of size $n=2$. Other parameters were $\nu=3$, $\alpha=0.27$ and $\beta=0.8$. A realisation was classified as a large outbreak if it led to a total of at least $15$ infected households.}\label{fig-identifiers}
\end{figure}

\boldtext{In our analysis we wish to be able to state that an outbreak has been `observed' in a particular neighbourhood. However, a chain of infection can quickly fizzle out, or continue to become a large, observable, outbreak. So detecting a small number of cases in a neighbourhood does not necessarily mean we are observing an outbreak there. Therefore we say that an outbreak is `observed' in neighbourhood when there have been at least $6$ infected households there. We determined this threshold empirically, as follows. We used the Gillespie SSA to simulate $10,000$ realisations of a single neighbourhood model composed of households of size $2$. A realisation was stopped if the number of infections reached zero or $t=100$ days. We classified realisations as a `large outbreak' if a total of $15$ or more households were infected over the course of the infection chain, and an `insignificant outbreak' otherwise.  Figure \ref{fig-identifiers} shows that a large outbreak occurred in approximately $40$\% of all realisations. However, if we condition on the number of infected households reaching at least $6$, then a large outbreak occurred in over $90$\% of realisations.}  

\boldtext{To examine the sequence of neighbourhoods in which an outbreak is observed, we seeded a large number of trials of the Gillespie SSA with a single infected individual in a random neighbourhood and recorded the order in which neighbourhoods reached the threshold of $6$ infected households. Trials in which the outbreak did not spread to every neighbourhood were removed before the next stage of the analysis.} 

\boldtext{There are a total of $720$ possible sequences of $6$ neighbourhoods. We assigned each sequence an arbitrary index and found the proportion of `outbreak' trials in which the infection was observed to spread between neighbourhoods in that sequence. We visualised these outcomes as a scatter plot showing the sequence proportion, or frequency, for each arbitrary sequence index. We then used the BIRCH clustering algorithm from the sklearn.cluster library (\cite{birch}) to group the neighbourhood sequences into three clusters according to the frequency with which the sequence was observed (see Figure \ref{fig-all-seq}). We used the boundaries of these clusters to classify sequences as relatively common, intermediate or rare. For example, in Figure \ref{fig-all-seq}, a sequence was classified as occurring relatively commonly if it was observed in at least $0.46$\% of trials. Having established this criterion for a sequence to be `common' we were able to find the total proportion of all trials accounted for by common sequences.}

\boldtext{All code used to produce the results in this study can be found at \url{https://github.com/ahb48/Neighbourhoods_and_households}.}

\section{Results}\label{sec-results}
\subsection{How neighbourhood localisation affects $R_*$}\label{subsec:neigh_loc_Rstar}
We begin by investigating how neighbourhood localisation impacts the household reproduction number in a population structured into two neighbourhoods with different household compositions.

\begin{figure}
            \centering
            \includegraphics[width=0.5\textwidth]{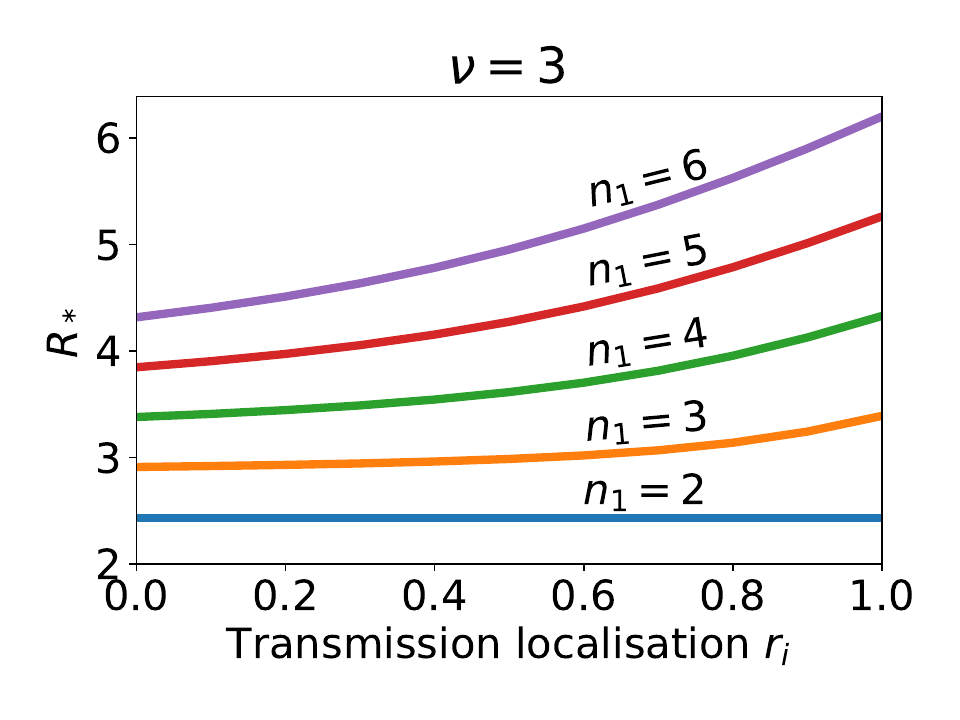}
    \caption{The household reproduction number for a population structured into two neighbourhoods as the localisation of transmission outside of households varies from proportional mixing to complete isolation. Each curve is for a different household size in neighbourhood $1$ ($n_2=2$ is fixed). $n_1=2,3,4,5,6$ corresponds to the blue, orange, green, red and purple lines respectively. Contact rate parameters are $\alpha=0.27$, $\beta=0.8$, which gives $R_* = 2.4$ when $n_1 = 2$.}\label{fig-beta}
\end{figure}

Figure \ref{fig-beta} shows how $R_*$ changes when the localisation of contact outside of households varies from proportional mixing between neighbourhoods to complete isolation. We observe that, when household size differs between the two neighbourhoods, increasing the localisation of transmission increases the expected number of infections arising from a single infected household in the early stages of an outbreak. This effect occurs because increasing the localisation increases the intensity of contact in both the neighbourhoods, including the one with the larger household sizes. Some of the contacts, and thus infections, that would have occurred in the neighbourhood with smaller households are retained in the neighbourhood with larger households. These larger households will, on average, go on to produce more subsequent infections than the smaller households. This effect is more pronounced for larger differences between household sizes. If both neighbourhoods have equal household sizes there is no effect. We observe the same result for a range of $\nu$ values. For larger $\nu$, i.e. when within household transmission is more important, we observe larger increases in $R_*$ when the localisation of the neighbourhoods is increased.

\begin{figure}
\centering
\includegraphics[width=0.8\textwidth]{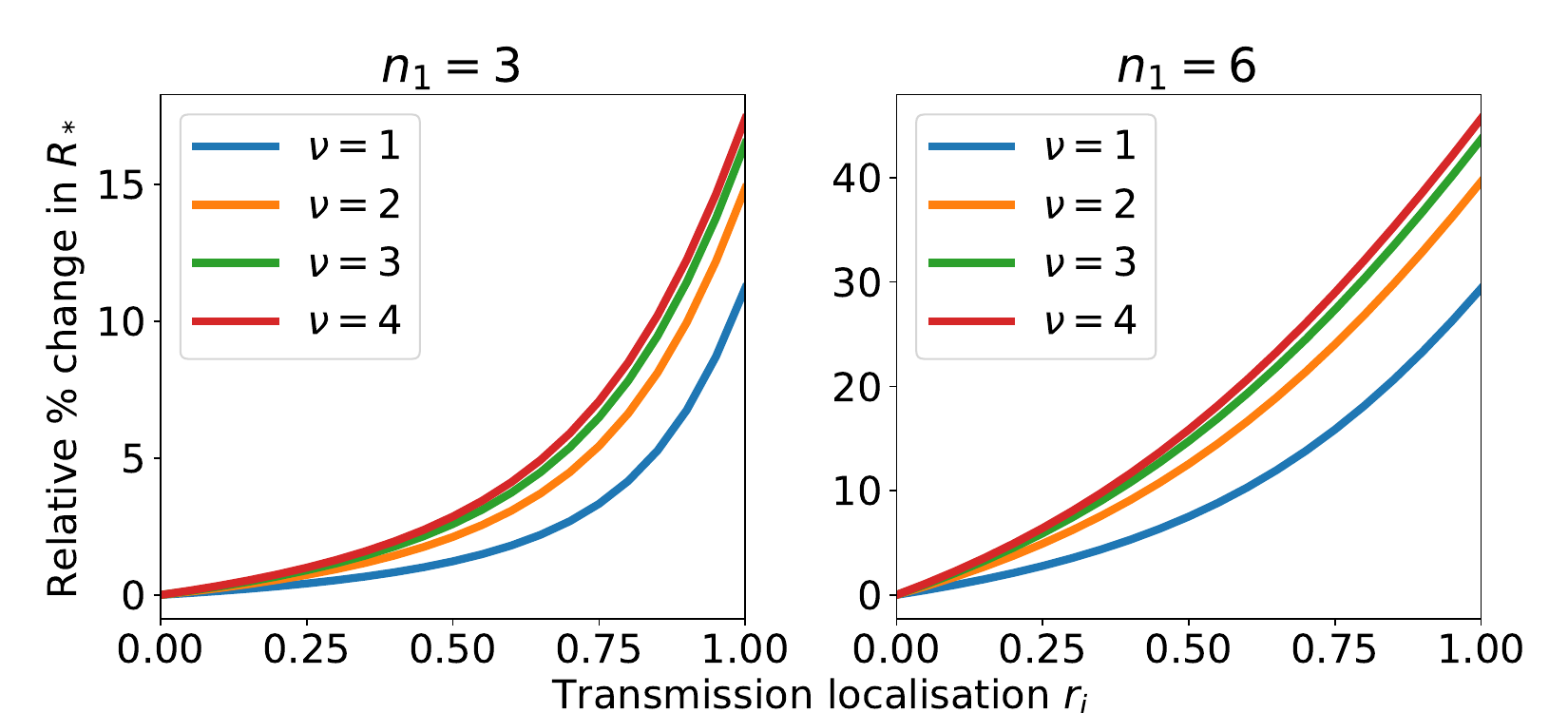}
\caption{The relative change in the value of the household reproduction number $R_*$ in a model with two neighbourhoods as localisation of transmission outside of households varies from proportional mixing between neighbourhoods to complete isolation. In both plots $n_2=2$ but household size in neighbourhood $1$ differs which modifies the baseline value of $R_*$ when $r_i=0$; see Figure \ref{fig-beta}. Each curve represents a different ratio of within household contacts to outside of household contacts. $\nu=1,2,3,4$ corresponds to the blue, orange, green and red lines respectively.}\label{fig-Rstar}
\end{figure}

Figure \ref{fig-beta} shows the absolute value of $R_*$. \boldtext{The parameterisation is such that $R_* = 2.4$ when $n_1 = n_2=2$.} But increasing $n_1$ increases $R_*$ independently of transmission localisation. So, to account for the baseline impact of changing $n_1$, Figure \ref{fig-Rstar} shows the relative change in $R_*$ when neighbourhood $2$ has households fixed at size $2$ and transmission localisation is varied from proportional mixing to complete isolation.
When $n_1=3$, the relative change in $R_*$ for proportional mixing versus isolated neighbourhoods ranges from $11.2$\% to $17.4$\% for the $\nu$ values we consider. In comparison, when $n_1=6$ the relative change in $R_*$ ranges from $29.4$\% to $45.7$\%. The relative change in $R_*$ is amplified by larger values of $\nu$ and $n_1$. 

We observe that weaker localisation of transmission outside of households can significantly reduce the epidemic risk and generation size early in the epidemic, even when the difference in household sizes between the two neighbourhoods is only one.

\subsection{How household size and neighbourhood localisation affects individual infection risk}\label{subsec:neigh_loc_ind_risk}
\begin{figure}
    \centering
    \begin{subfigure}[b]{0.5\textwidth}
            \centering
            \includegraphics[width=1\textwidth]{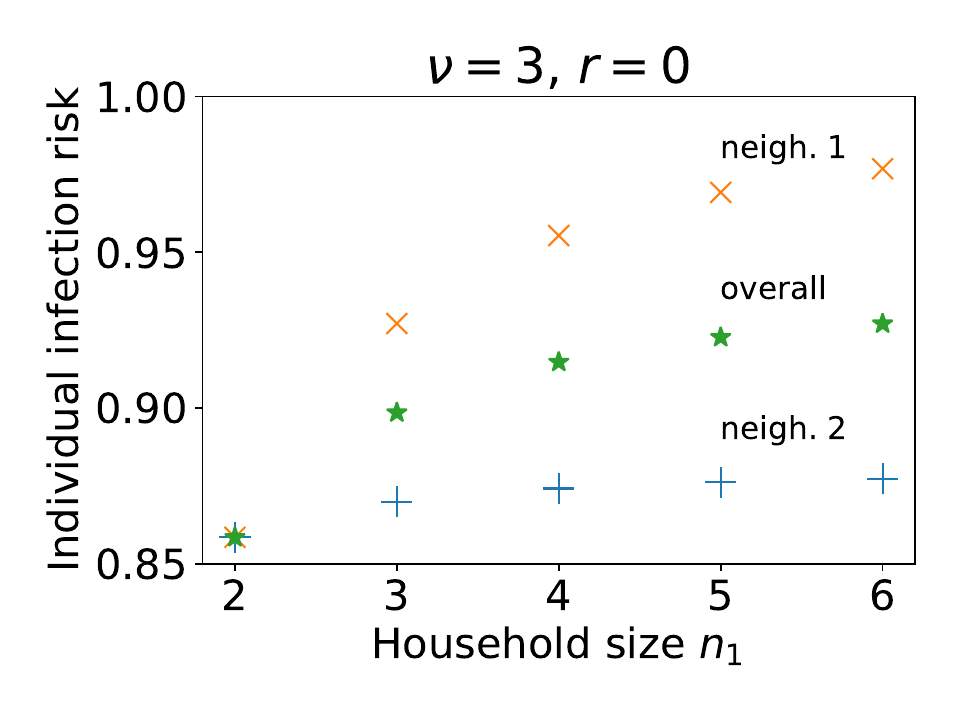}
            \caption{}
             \label{sfig:a}
    \end{subfigure}%
    \begin{subfigure}[b]{0.5\textwidth}
            \centering
            \includegraphics[width=1\textwidth]{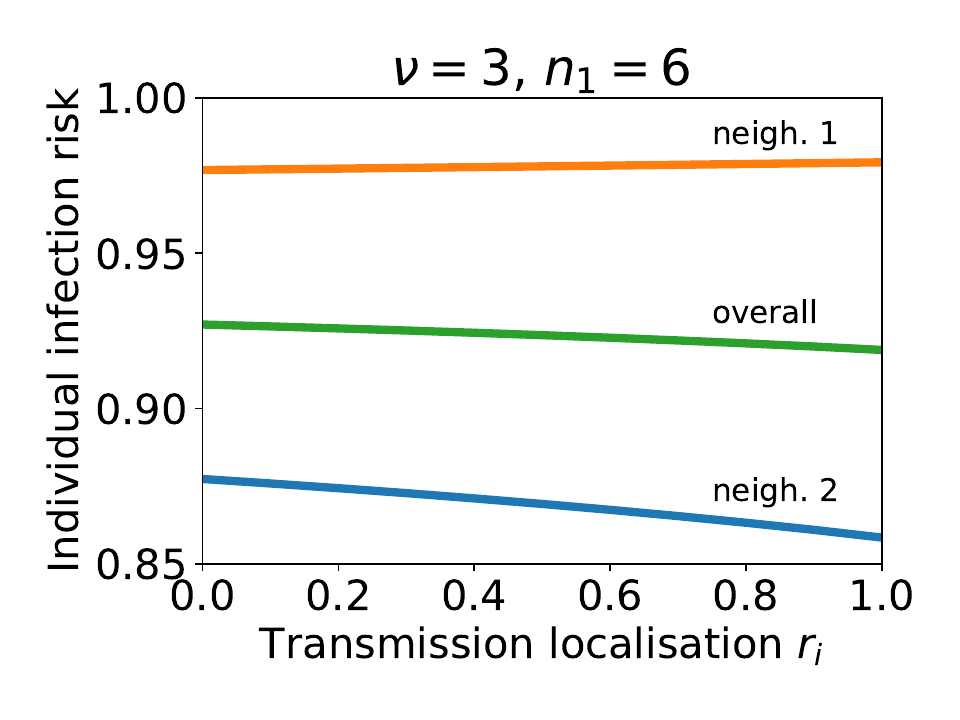}
            \caption{}
             \label{sfig:b}
    \end{subfigure}
    \caption{Individual infection risk over entire outbreak when neighbourhoods have different household sizes and the degree of localisation of neighbourhood transmission is varied. (\subref{sfig:a}) Household size in neighbourhood $1$ is varied from $n_1 = 2$ to $6$. Household size in neighbourhood 2, $n_2=2$. Individual infection risks of individuals from neighbourhood $1$, $2$ and overall are shown as blue crosses, orange pluses and green stars respectively. Infection risk is defined as the probability an individual has been infected (and is recovered) by the time the outbreak ends. (\subref{sfig:b}) Households sizes are fixed at $n_1=6$ and $n_2=2$. The localisation of neighbourhood contacts $r_i$ varies from proportional mixing to complete isolation. Infection risk of individuals from neighbourhoods $1$, $2$ and overall are displayed as blue, orange and green curves respectively.}\label{fig-risk}
\end{figure}

\boldtext{We consider the model with two neighbourhoods. Households in neighbourhood $2$ are of size $2$ and we examine how the infection risk depends on the household size in neighbourhood $1$. 
Figure \ref{fig-risk}\subref{sfig:a} shows the individual infection risk for someone from neighbourhood $2$ (households of size $2$), neighbourhood $1$ (households of size $2$ to $6$) and for someone chosen at random from the population. The localisation} corresponds to proportional mixing between neighbourhoods $(r_i = 0)$. The impact of household size on infection risk is localised. Individual infection risk in neighbourhood $1$ increases from $0.86$ to $0.98$ as household size increases from $2$ to $6$ whereas the increase in individual risk in neighbourhood $2$ is very small.

We also consider how the degree of localisation of contacts outside of the household impacts individual infection risk. Figure \ref{fig-risk}\subref{sfig:b} reveals that increasing the neighbourhood localisation of contacts decreases individual infection risk in the neighbourhoods with smaller households but has an insignificant impact on infection risk in the neighbourhoods with larger households.

We conclude that the impact of the household demography of a neighbourhood on individual infection risk over the entire course of the outbreak is mostly limited to the neighbourhood itself.
This contrasts with our previous observation in Figure \ref{fig-Rstar} that increased neighbourhood localisation of contacts outside of the household increases the epidemic risk when there is a difference in household size between the neighbourhoods.

\subsection{Outbreak probability and surveillance}\label{subsec:outprob_surv}
In this section we investigate how the probability of an outbreak in the two neighbourhoods model depends on neighbourhood household sizes and $\nu$. We further explore how outbreak probability estimates are affected by making different assumptions about the population structure. Simulation trajectories can show stuttering chains of transmission that die out within a few generations, or sustained periods of transmission with exponential growth patterns. We classify trajectories as representing an `outbreak' if there have been at least $6$ infected households (see Subsection \ref{subsec:methods_6neigh} for further details), and `no outbreak' otherwise. 

\begin{figure}
    \centering
    \begin{subfigure}[b]{0.5\textwidth}
            \centering
            \includegraphics[width=1\textwidth]{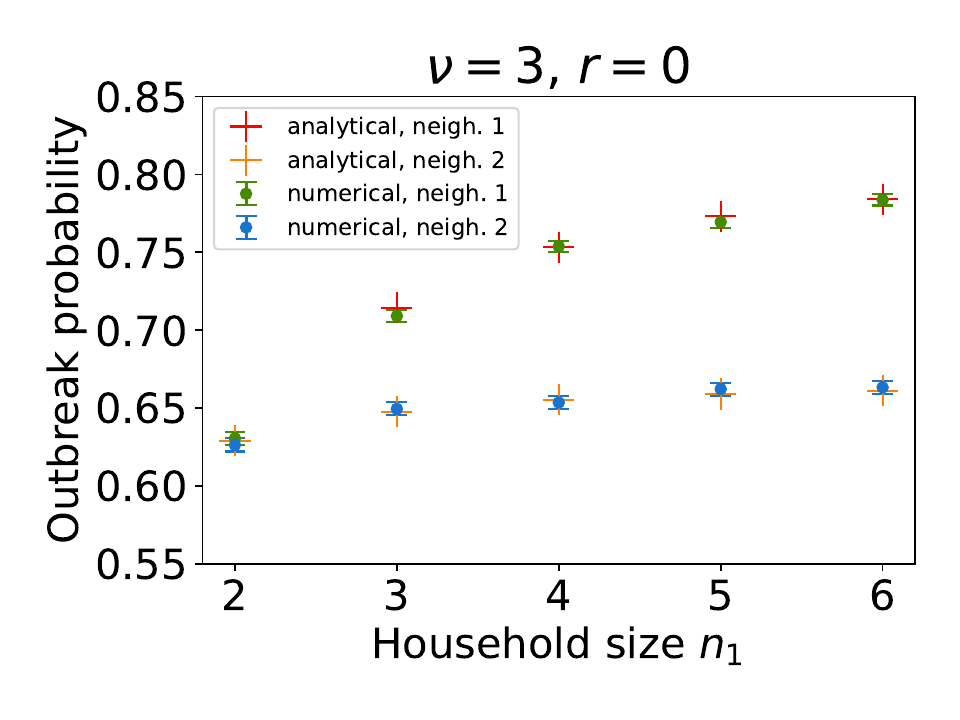}
            \caption{}
             \label{sfig-out-prob:a}
    \end{subfigure}%
    \begin{subfigure}[b]{0.5\textwidth}
            \centering
            \includegraphics[width=1\textwidth]{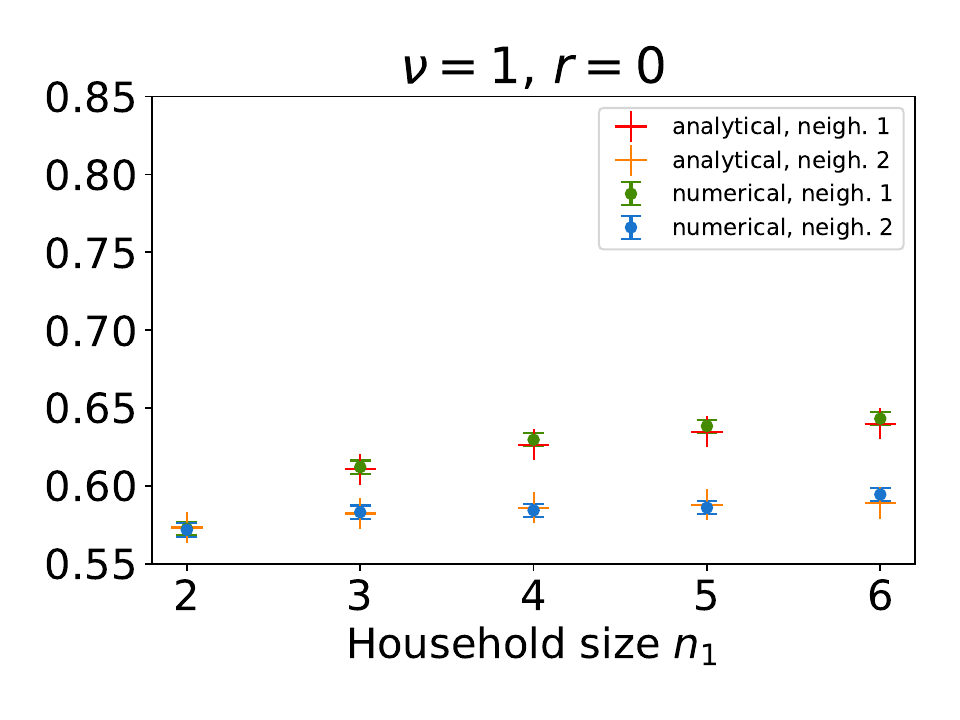}
            \caption{}
             \label{sfig-out-prob:b}
    \end{subfigure}
    \begin{subfigure}[b]{0.5\textwidth}
    \centering
    \includegraphics[width=1\textwidth]{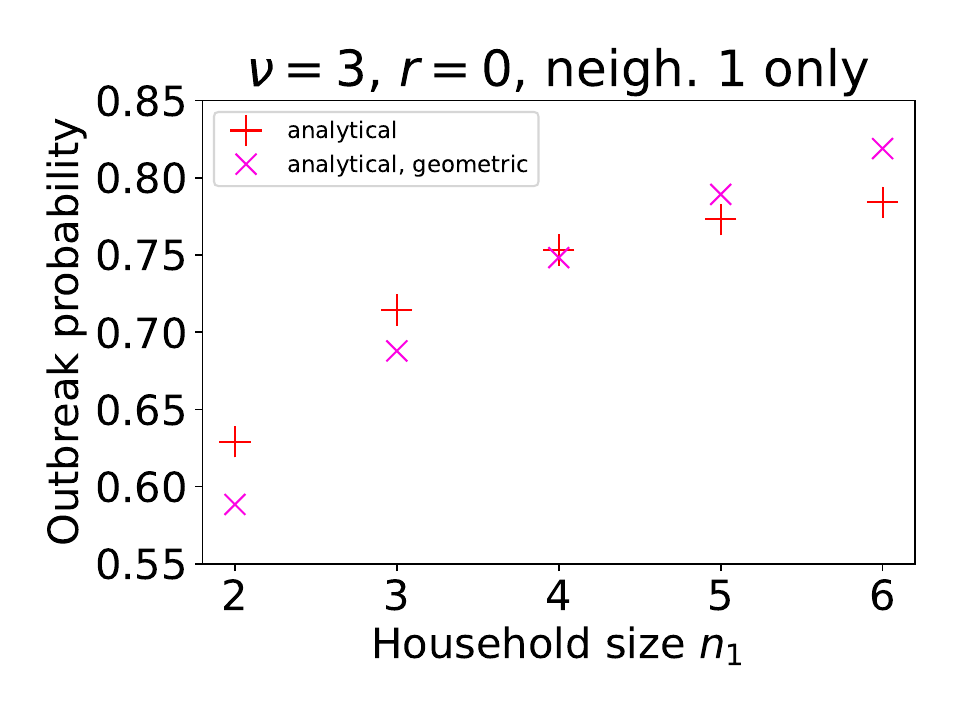}
    \caption{}
    \label{sfig-out-prob:c}
    \end{subfigure}%
    \begin{subfigure}[b]{0.5\textwidth}
    \centering
    \includegraphics[width=1\textwidth]{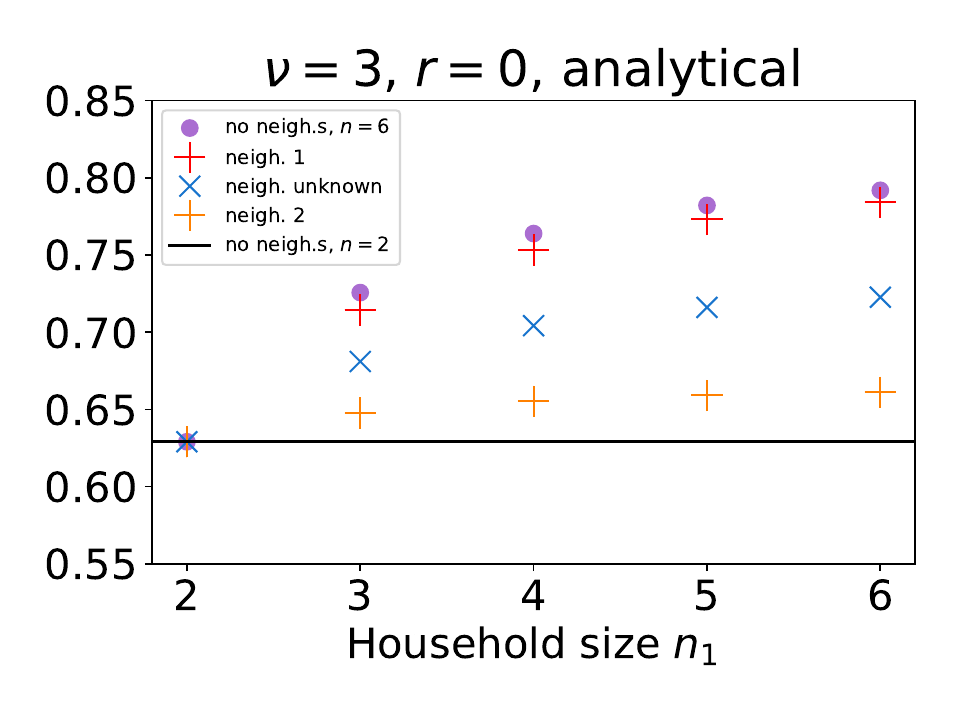}
    \caption{}
    \label{sfig-out-prob:d}
    \end{subfigure}
    \caption{\boldtext{(a) and (b):} The probability of an outbreak originating from a single infectious individual in neighbourhood $1$ (higher curve; green dots and red pluses) or a single infectious individual from neighbourhood $2$ (lower curve; blue dots and orange pluses), depending on the household size in neighbourhood $1$. Dots denote outbreak probabilities found via $50,000$ Gillespie SSA realisations \boldtext{($95\%$ CIs are indicated by bars in these cases)}; pluses denote the multi-type branching process approximation detailed in Subsection \ref{sec-prob-neigh}.
    Household size remains fixed at $n_2=2$ in neighbourhood $2$ and neighbourhood localisation corresponds to portionate mixing ($r_i = 0$). All other parameters are the same as in Table \ref{tab1} unless stated otherwise. (a) The ratio of within household contacts to outside of household contacts is set to $\nu=3$ and (b) $\nu=1$, and $\alpha, \beta$ are such that $\beta=\nu \alpha$ and $R_*=2.4$. (c) Probability of an outbreak originating from a single infectious individual in neighbourhood $1$ \boldtext{(calculated analytically)}, as in (a), with for comparison, naive branching process approximation based on a geometric offspring distribution shown as \boldtext{pink} crosses. (d) Probability of an outbreak calculated on the assumption that the initial infectious individual is: from neighbourhood $1$ (\boldtext{red pluses}); from neighbourhood $2$ (\boldtext{orange pluses}); from neighbourhood $1$ or $2$ with equal probability (\boldtext{blue crosses}); from a population of households of size $n_1$ with no neighbourhood structure (\boldtext{purple dots}). The solid black line denotes the outbreak probability given a single infected individual from a population of households of size $n_2=2$ (fixed).}\label{fig-out-prob}
\end{figure}

\boldtext{The analytic and numerical outbreak probability calculations are in good agreement.} Figure \ref{sfig-out-prob:a} shows that when the initial infected individual is from neighbourhood $1$, increasing the household size within this neighbourhood increases the probability of an outbreak. When $\nu=3$ and mixing is proportionate, for households of size $2$ the outbreak probability is $0.63$ and for households of size $6$ it is $0.78$ \boldtext{(green dots, red pluses)}. In comparison, the probability of an outbreak originating from an initial infected individual from neighbourhood $2$ is insensitive to the household composition of neighbourhood $1$ when $n_2=2$ remains fixed \boldtext{(blue dots, orange pluses)}. The increase in outbreak probability with neighbourhood household size is amplified by larger $\nu$. In Figure \ref{sfig-out-prob:b}, we see that smaller $\nu$ corresponds to outbreak probabilities in neighbourhood $1$ which are less responsive to changes in the household size in that neighbourhood \boldtext{(green dots, red pluses)}.

Figure \ref{sfig-out-prob:c} shows the outbreak probabilities again \boldtext{(red pluses)}, together with the outbreak probability calculated using a multi-type branching process approximation that naively assumes a geometric offspring distribution for households of any size \boldtext{(pink crosses)}. The probability generating functions for the geometric offspring distribution approximation are in the Supplementary information (Section $3$). For the specific parameter values used in this figure, the geometric offspring distribution assumption gives a good approximation to the true outbreak probability when $n_1 = 4$ (and $n_1=1$, not shown here), but diverges for other values of $n_1$.

\subsubsection{Surveillance}\label{subsec-surv}
\boldtext{One of the aims of an infectious disease surveillance system is to forecast the probability of a large outbreak when the first few cases are detected (\cite{arinaminpathy2009evolution,southall2023practical})}. 
Figure \ref{sfig-out-prob:d} explores the impact of different demographic assumptions when calculating the probability of a large outbreak on the basis of a small number of initial cases. As before, we assume there is initially a single infected individual in neighbourhood $1$, all households in neighbourhood $2$ are of size $2$, and all households in neighbourhood $1$ are of size $2$ to $6$. Therefore, the true outbreak probability is as in Figure \ref{sfig-out-prob:a} (\boldtext{red pluses}). Now we consider how this probability compares with outbreak probability calculations that do not fully account for the neighbourhood or household structure. If we neglect the neighbourhood structure entirely, and simply assume that all households in the community are of size $n_1$, we get a good approximation to the true outbreak probability (\boldtext{purple dots}). However, if we assume that all households in the community are of size $n_2$, we get a poor approximation to the true outbreak probability unless $n_1 = n_2$ (shown as black solid line). If we do account for the neighbourhood structure, but incorrectly assume that the initial infected individual is in neighbourhood $2$ then we get a similarly poor approximation to the true outbreak probability (\boldtext{orange pluses}). If we assume that the initial infected individual has an equal probability of being in each neighbourhood then we get a slightly better approximation (\boldtext{blue crosses}). This is simply the average of the outbreak probabilities with initial conditions of a single infected individual from neighbourhood $1$ or neighbourhood $2$ respectively. Therefore, we deduce that the demographic composition of the neighbourhood where the initial case occurs determines the outbreak probability, almost independently of the demography of the other neighbourhoods, even under proportional mixing.

\begin{figure}
    \centering
    \begin{subfigure}[b]{0.5\textwidth}
            \centering
            \includegraphics[width=1\textwidth]{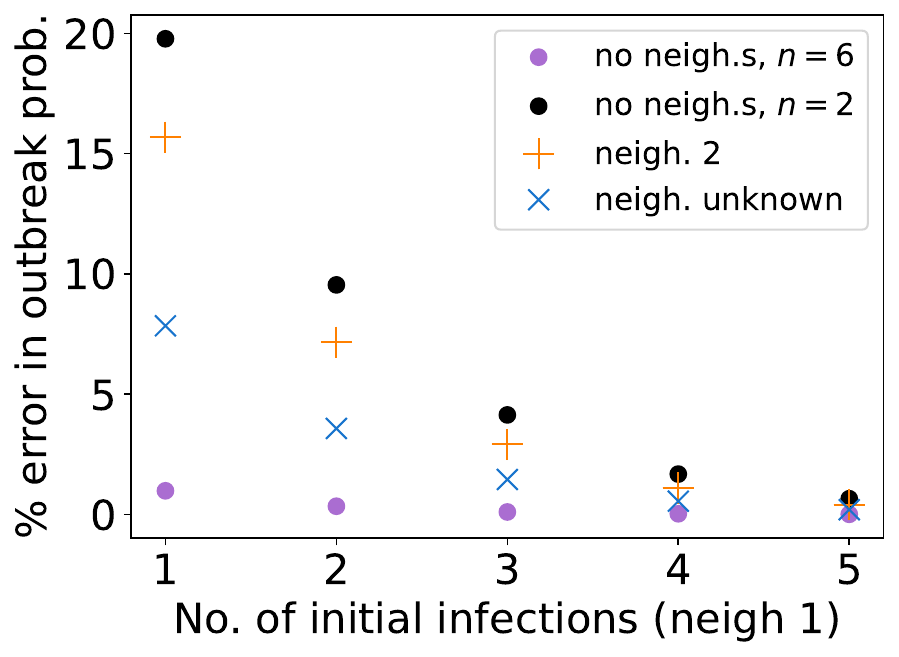}
            \caption{}
             \label{sfig:a}
    \end{subfigure}%
    \begin{subfigure}[b]{0.5\textwidth}
            \centering
            \includegraphics[width=1\textwidth]{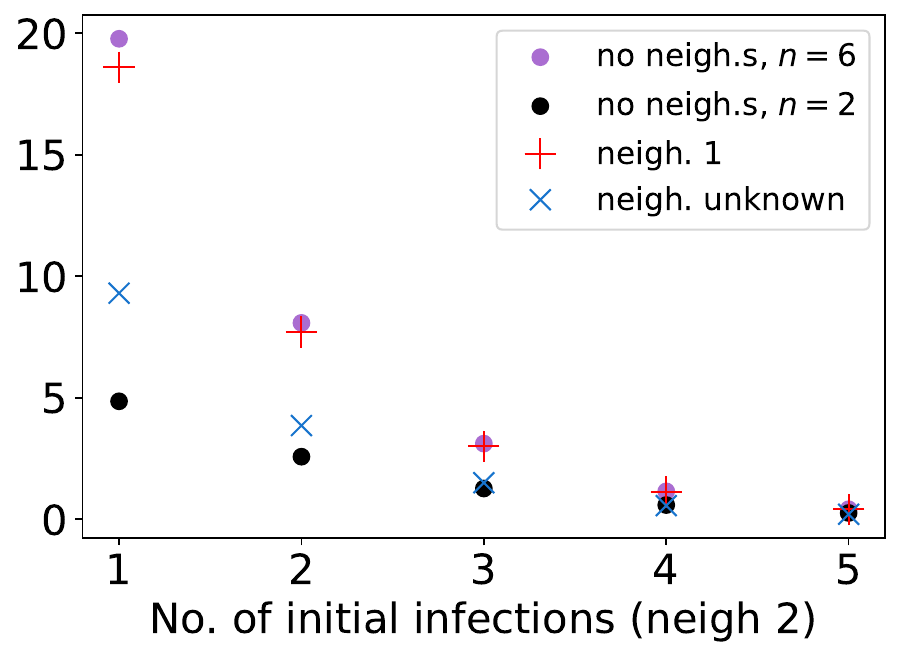}
            \caption{}
             \label{sfig:b}
    \end{subfigure}
    \caption{The relative error in the probability of an outbreak originating from $k=1,2,\ldots , 5$ initial infectious individuals (a) from neighbourhood $1$, $n_1=6$ and (b) from neighbourhood $2$, $n_2=2$ calculated under various assumptions. \boldtext{In all cases $\nu=3$ and $r=0$.} Assumptions: population of households size $n_1=6$ with no neighbourhood structure (blue pluses); population of households size $n_2=2$ with no neighbourhood structure (orange pluses); initial infected individuals all in the other neighbourhood (green crosses) and all in unknown neighbourhood (red stars).}\label{fig-surv}
\end{figure}

In Figure \ref{fig-surv} we show the error in the various outbreak probability calculations seen in the last column of Figure \ref{sfig-out-prob:d} with respect to the number of initial cases. In \boldtext{Figure \ref{sfig:a}}, when there is a single case within a household in neighbourhood $1$ ($n_1=6$) and it is erroneously assumed that the case originates in a population where all households are of size $n=2$, the relative percentage error in the outbreak probability is almost $20\%$ \boldtext{(black dots)}. The error is similar, Figure \ref{sfig:b}, if the initial case is from neighbourhood $2$ ($n_2=2$) but is erroneously assumed to be from a populations where all households are of size $n=6$ \boldtext{(purple dots)}. When the neighbourhood demography is modelled correctly but the initial case is attributed to the wrong neighbourhood, the relative error in the outbreak probability is slightly lower \boldtext{(see orange pluses in Figure \ref{sfig:a} and red pluses in Figure \ref{sfig:b})}. When the initial case is attributed to each neighbourhood with equal probability, the relative error is lower in both cases but still $7-10\%$ \boldtext{(blue crosses in Figures \ref{fig-surv})}. Finally, if the neighbourhood structure is neglected and all households are assumed to have the same size as that of the initial case then the error is close to $0$ when the initial case is from neighbourhood $1$ \boldtext{(Figure \ref{sfig:a}, purple dots)} and close to $5\%$ when it is from neighbourhood $2$ \boldtext{(Figure \ref{sfig:b}, black dots)}. This is due to the additional force of infection associated with larger households in neighbourhood $1$.

As the number of initial infected households increases, the error in each outbreak probability approximation decays exponentially towards $0$. After a chain of $4$ initial cases the errors are negligible. This is because the probability of a major outbreak is much higher (see Figure \ref{fig-identifiers}) for all demographic assumptions, reducing the relative scope for error.

\subsection{More complex neighbourhood structures model} \label{sec:6neigh}
\subsubsection{Neighbourhood in which an outbreak is first observed}
\begin{figure}
    \centering            
    \begin{subfigure}[b]{0.5\textwidth}
            \centering
            \includegraphics[width=1\textwidth]{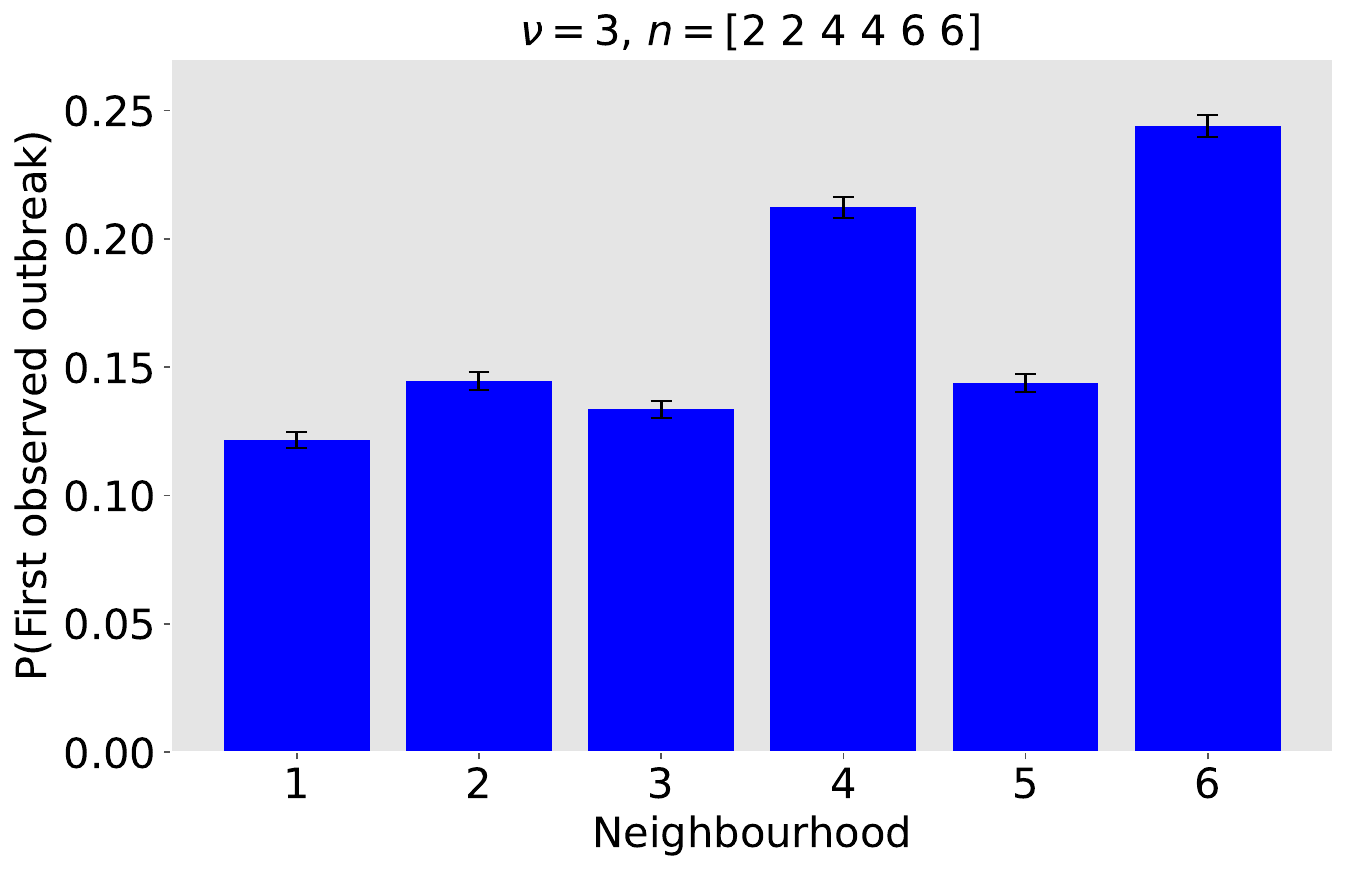}
            \caption{}
             \label{sfig-obs-risk:a}
    \end{subfigure}%
    \begin{subfigure}[b]{0.5\textwidth}
    \centering
    \includegraphics[width=1\textwidth]{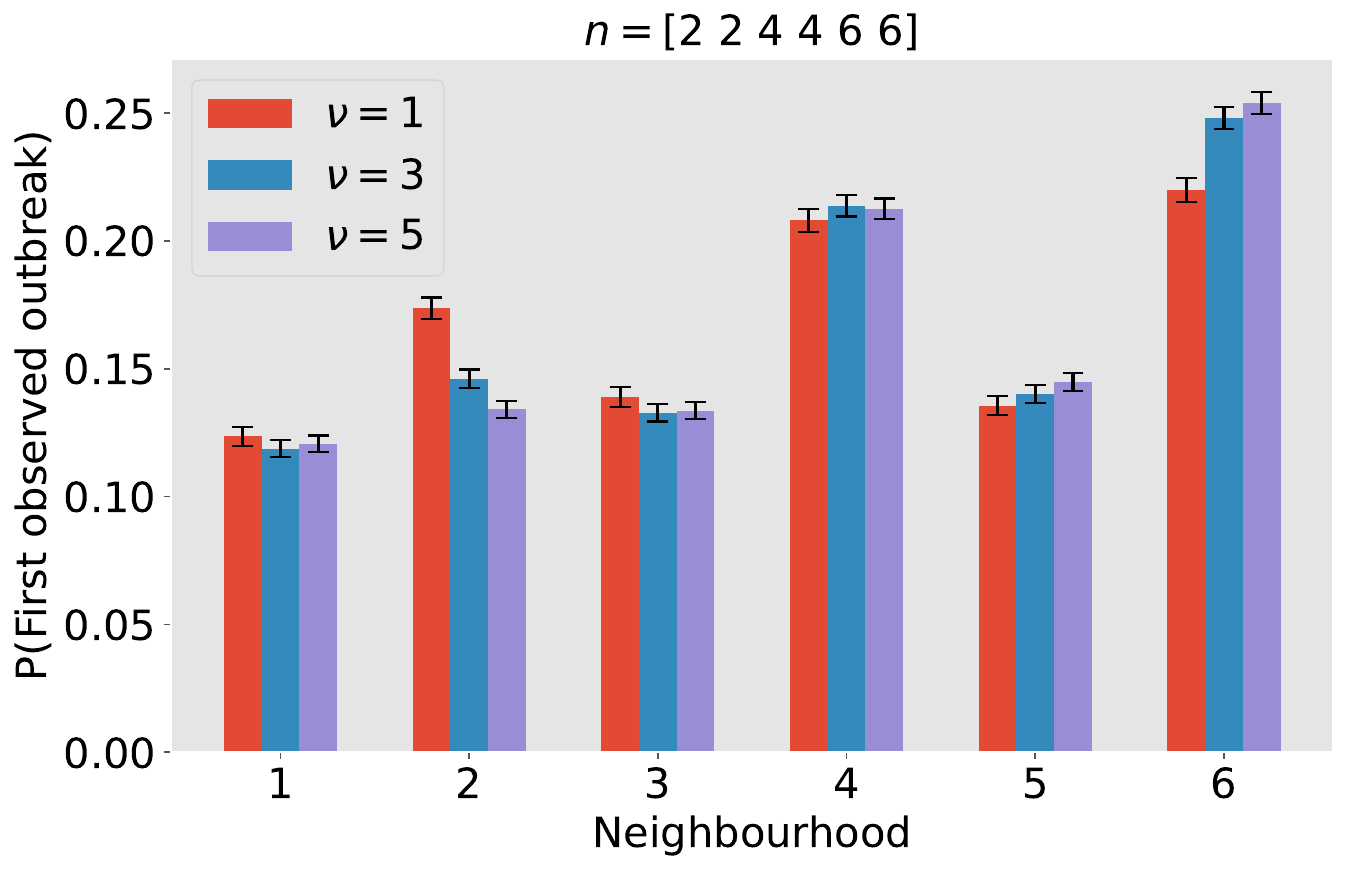}
    \caption{}
    \label{sfig-obs-risk:b}
    \end{subfigure}
    \caption{Probability that an outbreak is first observed in a given neighbourhood. The model has six neighbourhoods. Neighbourhoods $1$ and $2$ have households of size $2$, neighbourhoods $3$ and $4$ have households of size $4$, neighbourhoods $5$ and $6$ have households of size $6$. Neighbourhoods $1$, $3$ and $5$ have weak localisation of contact outside of households. Neighbourhoods $2$, $4$ and $6$ have intermediate localisation. An outbreak is `observed' in a neighbourhood when there have been at least $6$ infected households in that neighbourhood. Outbreak observation probabilities were calculated from $500,000$ Gillespie SSA realisations of the model. For each realisation, the localisation parameters $r_i$ were assigned randomly from the distribution $U[0,0.1]$ (neighbourhoods $1$, $3$, $5$) or $U[0.4,0.5]$ (neighbourhoods $2$, $4$, $6$) and the initial condition introduced a single infected individual into a randomly selected neighbourhood. \boldtext{The $95\%$ confidence intervals are represented as black lines.} (a) $\nu=3$ (b) $\nu=1,3,5$ in red, blue and purple respectively. 
    }\label{fig-obs-risk} 
\end{figure}

Figure \ref{fig-obs-risk} shows the probability that an outbreak is first observed in neighbourhood $j=1,2,\ldots , 6$. The model set up is described \boldtext{in Subsection \ref{subsec:methods_6neigh}.} The results are based on $500,000$ Gillespie SSA realisations of the model with an initial condition that seeds one infected individual into a randomly selected neighbourhood. So all neighbourhoods have the same probability of being the source of an outbreak. Figures \ref{sfig-obs-risk:a} and \ref{sfig-obs-risk:b} show the probability that an outbreak is first observed in each neighbourhood for the ratio of within household contacts to outside of household contacts, $\nu=3$ alone and $\nu=1,3,5$ overlaid for comparison. Outbreaks are more likely to be observed first in neighbourhoods where household sizes are larger, and where contact outside of the household is highly localised.  

The probabilities that the outbreak was first observed in the same neighbourhood as it was seeded are $0.5$, $0.52$ and $0.52$ for the more localised neighbourhoods ($2$, $4$ and $6$), almost independent of household size. Whereas for the less localised neighbourhoods ($1$, $3$ and $5$) the corresponding probabilities are $0.33$, $0.39$ and $0.41$, distinctly dependent on household size. Strong localisation increases the probability that an outbreak is first observed in the same neighbourhood as it was seeded because there are fewer opportunities for the chain of infection to `escape' into other neighbourhoods. When localisation is weaker, household size plays a more important role because rapid transmission in larger households amplifies transmission in the local neighbourhood. 

\boldtext{In Figure \ref{sfig-obs-risk:b}, we explore the impact that the ratio of within household contacts to outside of household contacts $\nu$ has on the neighbourhood in which an outbreak is first observed. The height of the red, blue and purple bars correspond to the neighbourhood probabilities when $\nu=1$, $3$, and $5$ respectively.} For a weaker contact rate within the household (verses contact outside of households i.e. low $\nu$), the amplification effect of large households is smaller, which reduces the probability that an outbreak is first observed in a neighbourhood composed of large households. This effect is most clearly seen in the neighbourhoods that have the smallest and largest household sizes and more localised contacts ($2$ and $6$).

\subsubsection{Neighbourhood infection sequences}
\begin{figure}
    \centering
    \includegraphics[width=0.8\textwidth]{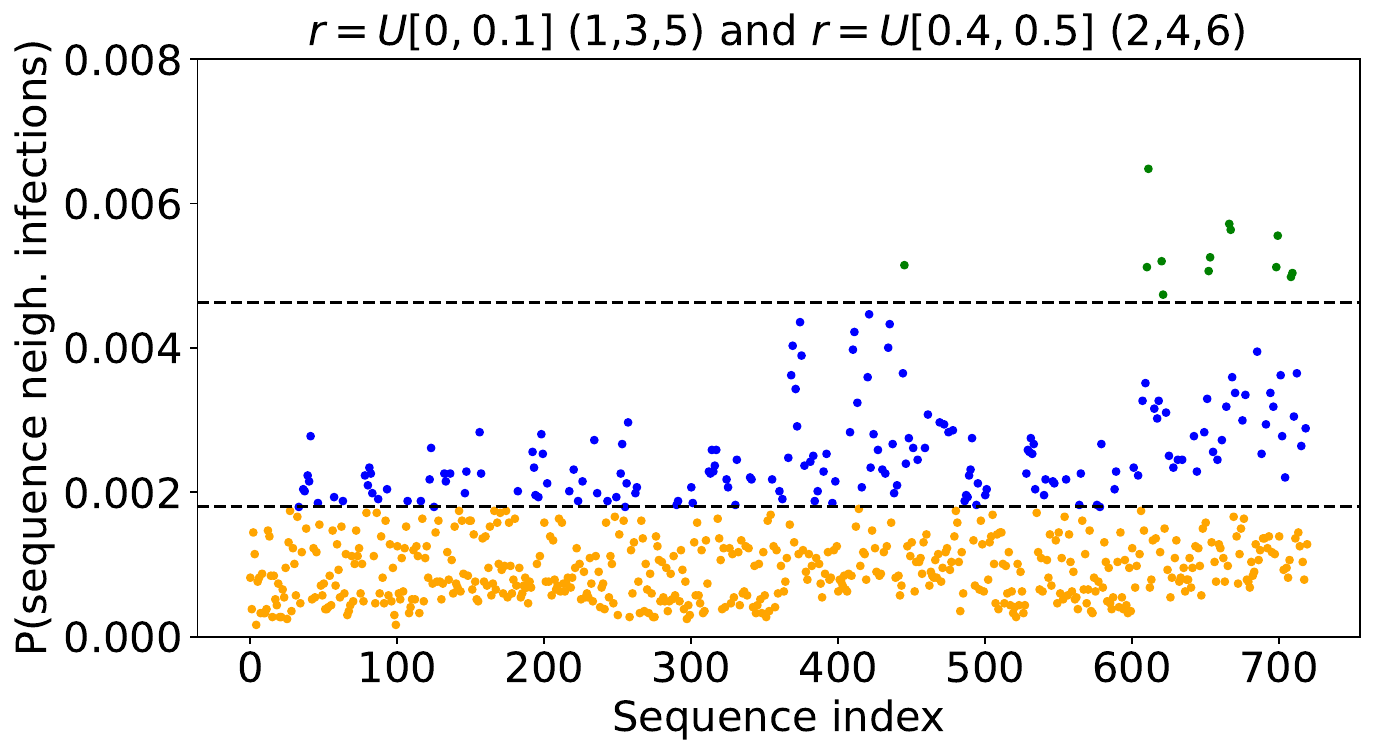}
    \caption{\boldtext{Probability an outbreak is observed in given sequences of neighbourhoods when neighbourhoods are grouped according to both household size and degree of localisation. There are $720$ possible sequences of $6$ neighbourhoods. Each point in the scatter plot corresponds to a unique sequence and shows the proportion of outbreaks in which the outbreak was observed in that sequence of neighbourhoods. \boldtext{The sequence index numbers are arbitrary. A table detailing the sequences corresponding to each index can be found at \url{https://github.com/ahb48/Neighbourhoods_and_households}.} A total of $50,000$ Gillespie SSA trials were computed. Those that did not result in \boldtext{large} outbreaks were discarded, leaving $36,724$ trials. Initially a single individual was infected in a randomly chosen neighbourhood. $\nu=3$ and $r$ was assigned a value from the distribution $U[0,0.1]$ (neighbourhoods $1$,$3$,$5$) or $U[0.4,0.5]$ (neighbourhoods $2$,$4$,$6$). The points above the first dashed line (coloured blue) correspond to sequences that occurred in $\geq 0.18\%$ of the outbreaks. Those above the second dashed line (coloured green) occurred in $\geq 0.46\%$ of outbreaks. Clustering was performed using the BIRCH algorithm in the `sklearn.cluster' library.}}
    \label{fig-all-seq}
\end{figure}

\begin{figure}
    \centering
    \includegraphics[width=0.9\textwidth]{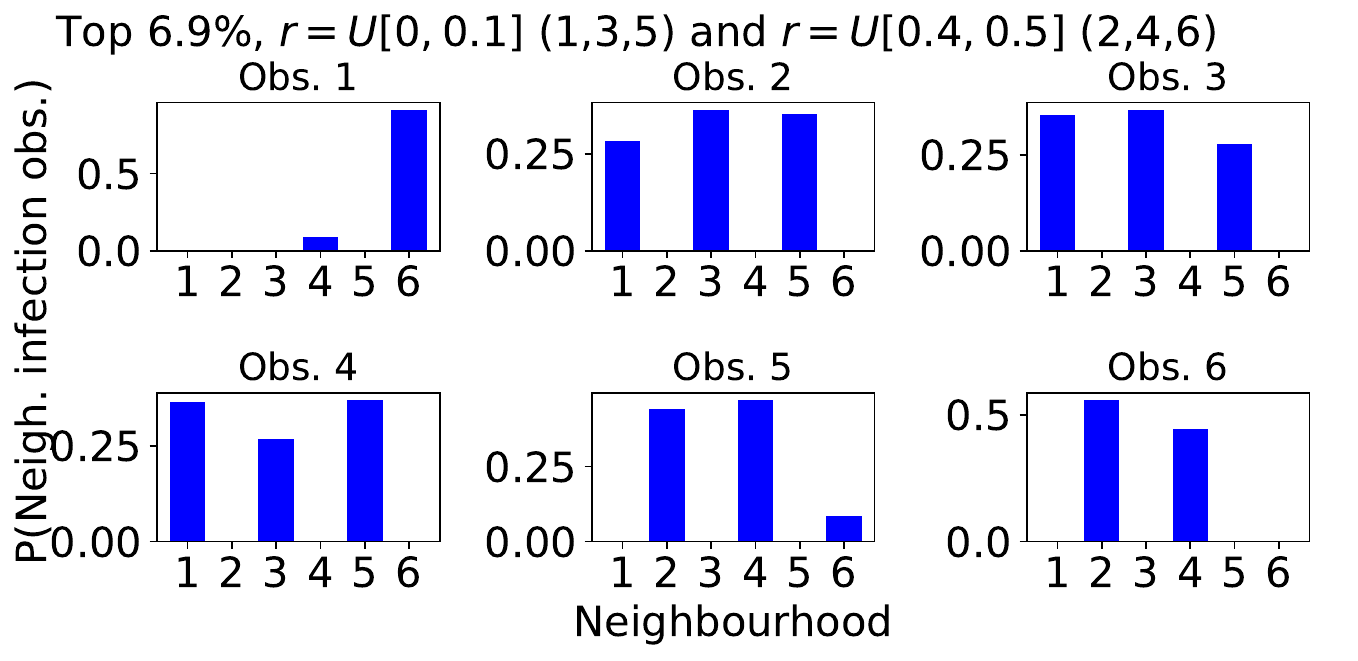}
    \caption{\boldtext{Typical sequence in which an outbreak is observed in different neighbourhoods when neighbourhoods are grouped according to both household size and degree of localisation. The model output is the same as in Figure \ref{fig-all-seq} but limited to the most common neighbourhood sequences accounting for a total of $6.9\%$ of trials. The top left panel shows the probability that each neighbourhood is the first in which the outbreak is observed. Subsequent panels show the probabilities that each neighbourhood is the $2$\textsuperscript{nd}, $3$\textsuperscript{rd}, ..., $6$\textsuperscript{th} in which an outbreak is observed.} $\nu=3$ and $r$ takes value from either the distribution $U[0,0.1]$ (neighbourhoods $1$,$3$,$5$) or $U[0.4,0.5]$ (neighbourhoods $2$,$4$,$6$).}
    \label{fig-high-seq}
\end{figure}

\boldtext{A large outbreak spreads through the $6$ neighbourhoods in one of $720$ possible sequences. Figure \ref{fig-all-seq} shows the frequency with which each of these neighbourhood sequences was recorded in $36,724$ trials when, as detailed in Subsection \ref{subsec:methods_6neigh}, the model was configured such that there were strongly localised neighbourhoods with small, medium and large households, and weakly localised neighbourhoods with small, medium and large households. The BIRCH clustering algorithm was used to group neighbourhood sequences into three groups according to the frequency with which they were recorded. In Figure \ref{fig-all-seq} the most common sequences are coloured green. Each of these $13$ sequences accounts for at least $0.46$\% of trails (upper dashed line), and together they account for $6.9$\% of all trials. For comparison, if all sequences were equally likely, each one would account for $0.14$\% of trials and any $13$ sequences would account for $1.8$\% of all trials. The sequences coloured blue in Figure \ref{fig-all-seq} each account for $0.18$-$0.45\%$ of trials. Those coloured orange each account for less than $0.18\%$ of trials, with many of them almost never observed. Focusing on the $13$ most common sequences that account for $6.9\%$ of trials, Figure \ref{fig-high-seq} shows the probability that each neighbourhood is the $1\textsuperscript{st}$, $2\textsuperscript{nd}$, $3\textsuperscript{rd}$, $\ldots$, $6\textsuperscript{th}$ in which the outbreak is observed. We see that} the outbreaks are always observed first in neighbourhoods $4$ or $6$. These are the neighbourhoods with households of size $4$ and $6$ with strong localisation of contact outside of the household $r \sim U[0.4,0.5]$. The outbreaks are then usually observed in neighbourhoods $1$, $3$ and $5$. These are the neighbourhoods of size $2$, $4$ and $6$ with weak localisation. Finally, outbreaks are usually observed last in neighbourhoods $2$ and $4$ once again where localisation is stronger. So, in summary, outbreaks tend to be observed first in a strongly localised larger household size neighbourhood because they can gather initial momentum there. They then spread through the less localised (i.e. more connected) neighbourhoods before eventually reaching the remaining highly localised (weakly connected) neighbourhoods. 

We observe similar results for a range of values for the ratio of within household contacts to outside of household contacts $\nu$; see Supplementary information Figures ($1$ and $2$). The pattern in the most frequently observed sequences of neighbourhood infections becomes less prominent as $\nu$ is decreased since household size becomes less important.

\begin{figure}
    \centering
    \includegraphics[width=0.8\textwidth]{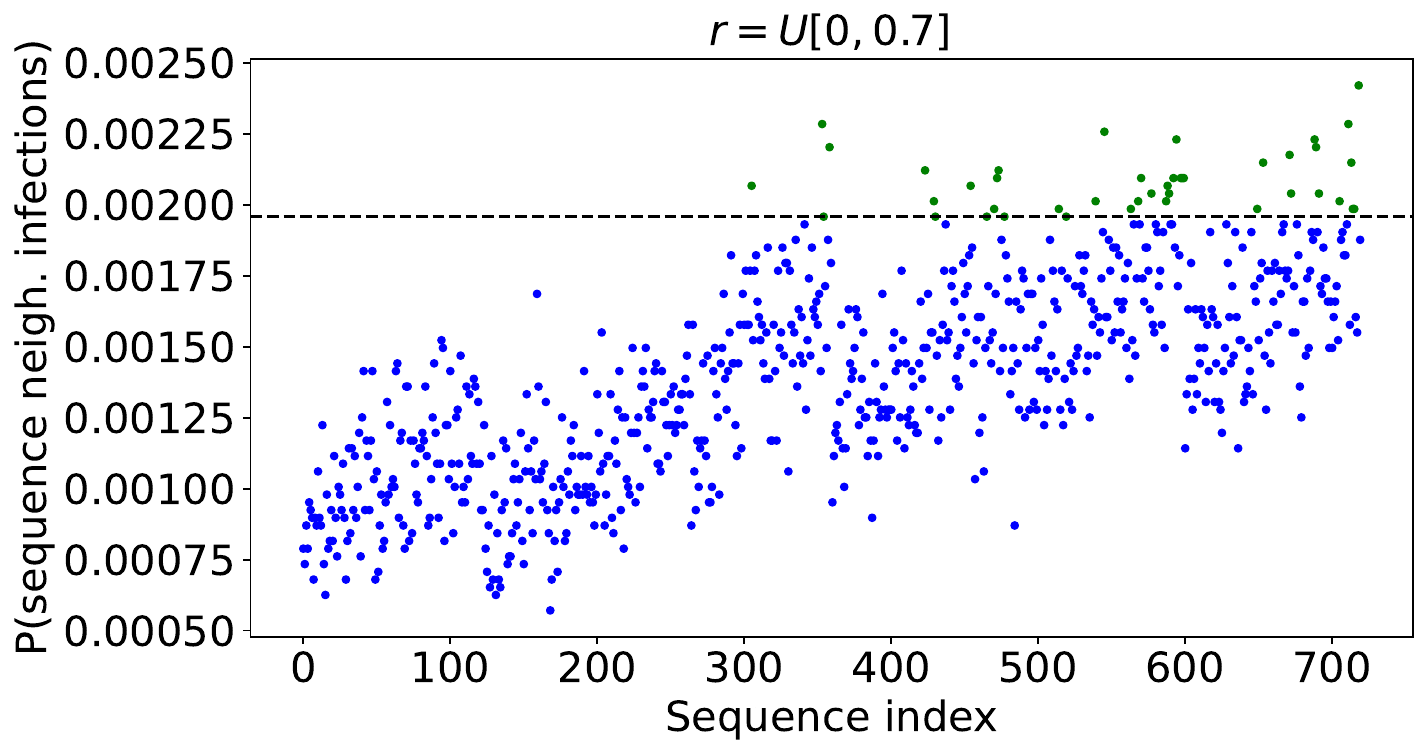}
    \caption{Probability an outbreak is observed in given sequences of neighbourhoods when neighbourhoods are grouped by household size but not the degree of localisation. There are $720$ possible sequences of $6$ neighbourhoods. Each point in the scatter plot corresponds to a unique sequence and shows the proportion of outbreaks in which the outbreak was observed in that sequence of neighbourhoods. \boldtext{The sequence index numbers are arbitrary. A table detailing the sequences corresponding to each index can be found at \url{https://github.com/ahb48/Neighbourhoods_and_households}.} Calculated from $50,000$ Gillespie SSA trials with those that did not result in \boldtext{large} outbreaks discarded, leaves $36,759$ trials. Initially a single individual was infected in a randomly chosen neighbourhood. $\nu=3$ and $r$ is assigned a value from the distribution $U[0,0.7]$. The green points above the first dashed line correspond to the most common sequences that account for a total of $6.9$\% of all trials.}
    \label{fig:all_seq_ranr}
\end{figure}

\begin{figure}
    \centering
    \includegraphics[width=0.9\textwidth]{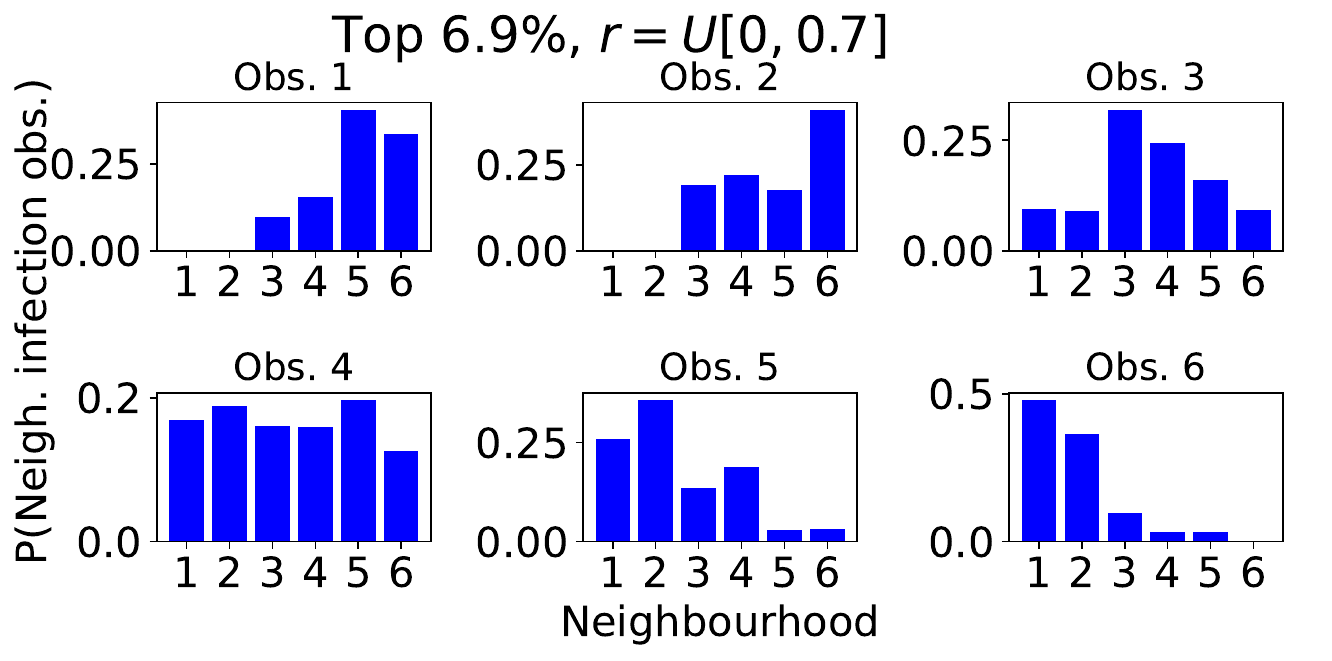}
    \caption{Typical sequence in which an outbreak is observed in different neighbourhoods when neighbourhoods are grouped by household size but not the degree of localisation. The model output is as in Figure \ref{fig:all_seq_ranr} but limited to the most common neighbourhood sequences accounting for a total of $6.9\%$ of trials. The top left panel shows the probability that each neighbourhood is the first in which the outbreak is observed. Subsequent panels show the probabilities that each neighbourhood is the $2$\textsuperscript{nd}, $3$\textsuperscript{rd}, ..., $6$\textsuperscript{th} in which an outbreak is observed. $\nu=3$ and $r$ takes value from the distribution $U[0,0.7]$.}
    \label{fig-high-seq2}
\end{figure}

\boldtext{We also examined the neighbourhood sequence in which an outbreak was observed when neighbourhoods were distinguished by household size; the localisation parameters were all assigned randomly from the same distribution. We ran $50,000$ trials and, as before, removed those that did not result in a large outbreak. This left $36,759$ trials. Figure \ref{fig:all_seq_ranr} shows the frequency with which each neighbourhood sequence occurred in these trials. The split between common and rare sequences is less distinct than in Figure \ref{fig-all-seq} and so, rather than clustering, we seek to compare with Figure \ref{fig-all-seq} by focusing on the set composed of the $39$ most common sequences that account for a total of $6.9\%$ trials in which large outbreaks occurred.}

In comparison to the model in which neighbourhoods are grouped by both household size and localisation, the pattern in the neighbourhood sequences in which the outbreak was observed is weaker; see Figure \ref{fig-high-seq2}. However, the feature of outbreaks being observed first in neighbourhoods with larger households and last in those with smaller households remains. 

Furthermore, the expected value of the localisation parameters $r$ for the 
$1\textsuperscript{st}$, $2\textsuperscript{nd}$, $3\textsuperscript{rd}$, $\ldots$, $6\textsuperscript{th}$ neighbourhoods observed in the full set of outbreaks was $0.38$, $0.3$, $0.3$, $0.32$, $0.36$ and $0.45$.
This once more demonstrates the pattern of outbreaks being observed first and last in more localised neighbourhoods, emerging in the less localised neighbourhoods in between.

\section{Discussion}\label{sec-disc}
In this paper, we have explored the intertwined role of households and neighbourhoods in the early stages of an epidemic. We harnessed fundamental theory of household epidemics developed in very general contexts to study the specific and tangible context of a metapopulation of households where contacts can occur both within households and outside of households at different rates. We focused on how several key quantities \boldtext{(neighbourhood household size, localisation, importance of contact within-households relative to contact in the wider community)} impacted the dynamics of an infectious disease spreading through this population.

We constructed the household reproduction number $R_*$ for the two neighbourhood model. We found that, when neighbourhoods are demographically characterised by different household sizes, greater localisation of community contact increases $R_*$. 
This is because more contacts from the larger household size neighbourhood occur with individuals from their own neighbourhood, leading to a larger number of subsequent chains of infection in the larger households. Larger infected households will on average produce more infections than smaller infected households. Thus, the overall expected number of infected households is larger for greater localisation of contacts.
$R_*$ is more sensitive to these parameters when the relative importance of within-household contact is higher or there is a bigger difference in the household sizes of the two neighbourhoods. 

We derived the analytic probability of a \boldtext{large} outbreak for the two neighbourhood model. We found that increasing the household size of a neighbourhood increases the probability that a single infected individual in that neighbourhood starts an outbreak, but has only a modest impact on the probability of an outbreak originating from a single case in the other neighbourhood.
Similarly, an individual's risk of infection was found to only be impacted by household size in their residential neighbourhood. Increasing neighbourhood localisation of contacts was shown to decrease individual infection risk in the smaller household size neighbourhood but had little impact on the larger household size neighbourhood.

We investigated the epidemiological dynamics in a model with six neighbourhoods using the Gillespie SSA. We found that population-wide outbreaks are more likely to be detected first in neighbourhoods with more localised community contact. We found considerable stochastic variation in the overall sequence of neighbourhoods in which outbreaks are detected. But, in general, outbreaks tended to be detected first in neighbourhoods with more localised community contact and large households, then in neighbourhoods with less localised contact throughout the middle stages, and finally in neighbourhoods with more localised contact but smaller household sizes. This pattern is amplified by increased relative importance of within-household contact.

We chose epidemiological parameters consistent with an acute respiratory infection such as influenza. We were unable to obtain reliable estimates for the within-household contact rates so \boldtext{we set $R_*=2.4$, consistent with \cite{black2013epidemiological}} and used this to find the community contact rate $\alpha$ and the within-household contact rate $\beta$ for our chosen values of the relative importance of within-household contact $\nu$. A productive avenue for future work may be to establish more reliable household contact rate estimates from data, or sample the parameter from an empirically motivated distribution as in \boldtext{\cite{black2013epidemiological}}.

Our work was inspired by several case studies and city planning documents which highlighted household size to be a key neighbourhood characteristic \boldtext{(\cite{kamata2010mongolia,ibitoye2017spatial})}.
When populations are structured \boldtext{as households clustered with similar household sizes (or share other demographic characteristics)}, our model may offer some useful insights into infectious disease surveillance and control strategies. We found that, when we wish to estimate the probability of a large outbreak on the basis of a small number of initial cases, household information for those initial infected individuals is more useful than knowing the demographic composition of the other neighbourhoods. \boldtext{Many countries are now looking to `smart growth' policies (\cite{kamata2010mongolia}) in order to accommodate growing populations in urban areas. These policies often involve building upwards rather than outwards which may increase neighbourhood and community population densities but lead to smaller households and lower family densities within individual houses. These changes in population density at different scales may act synergistically and have significant implications for infectious disease epidemiology.}

\boldtext{Our work uses a relatively simple model to allow some analytic tractability and provide fundamental insights into the epidemiological implications of multi-scale demographic structures. There are many ways in which this modelling framework can be extended to explore finer details of the demographic heterogeneities. But the increased complexity will make mathematical analysis and mechanistic interpretation more challenging and likely require agent-based simulation approaches.}  

\boldtext{For example, we fixed parameters in our model such as the local population size, within household contact rate and the recovery rate across all neighbourhoods. We also assumed that household size is homogeneous within any given neighbourhood. In reality, within-household contact rates are likely to vary between neighbourhoods, for instance due to different sanitation systems and building compositions. Neighbourhood population sizes are likely to vary, for example in the city centre versus the suburbs, and there is a distribution of household sizes within any neighbourhood. We modelled the neighbourhood localisation of contact by assuming a proportion of an individual's contacts outside of their own household are `reserved' for other individuals from the same neighbourhood. The remaining non-household contacts can occur at random with any individual from the entire population. However, many contacts outside of the home are likely to be structured around locations where people regularly congregate.} 

\boldtext{All of these heterogeneties can be brought into the model, at the expense of tractability. The general framework for modelling household epidemics has been extended to incorporate distributions of household sizes (\cite{black2013epidemiological,ross2015contact}), and other studies have introduced demographic features such as age-specific contact rates (\cite{geard2015effects, hilton2019incorporating}) and population aggregation in workplaces and schools \boldtext{(\cite{pellis2009threshold})}. In future work we plan to adapt our multiscale framework to incorporate some of these more complex demographic structures and tease apart their epidemiological impacts.}

\boldtext{In addition to the simplified demographic structures, our analysis assumes that surveillance is always perfect and all cases are detected immediately. In reality, many cases are undetected, there are reporting delays, and there is only partial information about the circumstances of individual cases. Hence future work will examine the impact of such partial information on outbreak predictions. One possible starting point may be recent work on the information content of cross-sectional versus cohort study sampling designs for fixed household sizes (\cite{kinyanjui2016information})}.

In conclusion, infectious disease epidemiology is shaped by demographic structures at several scales including households and neighbourhoods. Accounting for these structures can lead to a better understanding of epidemic risk and the patterns of epidemic spread and support more robust surveillance strategies.

\backmatter

\bmhead{Supplementary information}
All code used to produce the results in this piece of work can be found at \url{https://github.com/ahb48/Neighbourhoods_and_households}. Supplementary information can be found in a supporting document online.

\bmhead{Acknowledgments}
This work is supported by a scholarship from the EPSRC Centre for Doctoral Training in Statistical Applied Mathematics at Bath (SAMBa), under the project EP/S022945/1.
For the purpose of open access, the author has applied a Creative Commons Attribution (CC-BY) licence to any Author Accepted Manuscript version arising.
\boldtext{We thank the two referees for their useful feedback.}
No new data were created during the study. 

\bibliography{sn-bibliography}


\end{document}